\documentclass{article}

\PassOptionsToPackage{hyphens}{url}
\usepackage[preprint]{neurips_2026}
\usepackage{tabularx}
\usepackage{latexsym}
\usepackage[T1]{fontenc}
\usepackage[utf8]{inputenc}
\usepackage[table,dvipsnames]{xcolor}
\usepackage[colorlinks=true,linkcolor=black,citecolor=black,urlcolor=black]{hyperref}
\usepackage{url}
\usepackage{microtype}
\usepackage{inconsolata}
\usepackage{amsmath}
\usepackage{amssymb}
\usepackage{amsfonts}
\usepackage{graphicx}
\usepackage{booktabs}
\usepackage{colortbl}
\usepackage{subcaption}
\usepackage{enumitem}
\usepackage{placeins}
\usepackage{afterpage}
\usepackage{needspace}
\usepackage{multirow}
\usepackage{pifont}
\setcitestyle{round}
\makeatletter
\providecommand*\@LN[2]{}
\providecommand*\@LN@col[1]{}
\makeatother
\newcommand{\lhb}{\textsc{LoopsBench}}

\newcommand{\rev}[1]{#1}
\newcommand{\benchYes}{\textcolor{ForestGreen}{\ding{51}}}
\newcommand{\benchNo}{\textcolor{red}{\ding{55}}}
\newcommand{\benchPartial}{\textit{partial}}
\usepackage{tcolorbox}
\tcbuselibrary{skins,breakable,listings}
\definecolor{lhbPromptHead}{HTML}{4A6FA5}
\definecolor{lhbPromptBody}{HTML}{F2F5FB}
\definecolor{lhbPromptRule}{HTML}{C8D4E8}
\lstdefinestyle{lhbPrompt}{%
  basicstyle=\ttfamily\footnotesize,
  breaklines=true,
  breakindent=0pt,
  breakatwhitespace=false,
  columns=fullflexible,
  keepspaces=true,
  showstringspaces=false,
  upquote=true,
  literate={`}{{\textasciigrave}}1
}
\newtcblisting{promptbox}[2][]{%
  enhanced, breakable, sharp corners=south,
  colback=lhbPromptBody, colframe=lhbPromptHead,
  coltitle=white, fonttitle=\small\bfseries\sffamily,
  title={#2}, before skip=4pt, after skip=4pt,
  left=4pt, right=4pt, top=2pt, bottom=2pt,
  boxrule=0.5pt, titlerule=0pt,
  listing only, listing options={style=lhbPrompt}, #1
}
\newcommand{\parabf}[1]{\noindent\textbf{#1}~}
\newcommand{\authmark}[1]{\textsuperscript{\normalfont #1}}
\newcommand{\authornamefont}{\normalfont\small\bfseries}
\newcommand{\authorlink}[2]{\href{#1}{\textcolor{black}{{\authornamefont #2}}}}

\title{\lhb{}: From Harness Engineering to Loop Engineering in Coding Agent Evaluation}

\author{%
  \authorlink{https://openreview.net/profile?id=~Han_Li44}{Han Li}\authmark{1,\ensuremath{\ddagger},*}\quad
  \authorlink{https://openreview.net/profile?id=~Zhemin_Fang1}{Zhemin Fang}\authmark{2,*}\quad
  \authorlink{https://openreview.net/profile?id=~Rili_Feng1}{Rili Feng}\authmark{2,*}\quad
  \authorlink{https://openreview.net/profile?id=~Yingqi_Zhao2}{Yingqi Zhao}\authmark{4}\quad
  {\authornamefont Jiaheng Liu}\authmark{2}\quad
  \authorlink{https://openreview.net/profile?id=~Pengfei_Gao1}{Pengfei Gao}\authmark{1,\textdagger}
  \\
  \authorlink{https://openreview.net/profile?id=~He_Ye2}{He Ye}\authmark{3}\quad
  \authorlink{https://openreview.net/profile?id=~Dayi_Lin1}{Dayi Lin}\authmark{1}\quad
  \authorlink{https://openreview.net/profile?id=~Qingwei_Lin1}{Qingwei Lin}\authmark{1}\quad
  \authorlink{https://openreview.net/profile?id=~Saravan_Rajmohan3}{Saravan Rajmohan}\authmark{1}\quad
  \authorlink{https://openreview.net/profile?id=~Dongmei_Zhang2}{Dongmei Zhang}\authmark{1}
  \\
  \footnotesize\authmark{1}Microsoft\quad
  \authmark{2}Nanjing University\quad
  \authmark{3}University College London\quad
  \authmark{4}Shanghai Jiao Tong University
  \\
  \footnotesize\textbf{Project page:} \href{https://loopsbench.ai/}{\texttt{https://loopsbench.ai/}}
  \\
  \footnotesize\authmark{*}Equal contribution.\quad\authmark{\textdagger}Corresponding author.\quad\authmark{\ensuremath{\ddagger}}Work done during an internship at Microsoft.%
}

\begin{document}

\maketitle
\suppressfloats[t]

\begin{abstract}
Coding agent infrastructure is shifting from harness engineering toward loop engineering as coding agents are deployed for sustained long horizon software development. Existing benchmarks often center on localized tasks or end state outcomes, offering limited insight into sustained execution. We introduce \lhb{}, a long horizon benchmark for loop engineering in coding agent evaluation. Each task is a dependency DAG over separately testable development units with source evidenced prerequisite edges. \lhb{} comprises 112 tasks from authentic sources spanning 8 programming languages and 9 domains. Its flow aware runtime releases tests along the ready frontier and retains completed nodes as regression obligations. We evaluate frontier coding agents paired with widely used loop implementations. The strongest configuration, Opus-4.7 with Claude Code and outer continuation, resolves 25.00\% of tasks. \rev{Recorded plans recover only part of the source recovered prerequisite DAG,} and regression events remain visible across the evaluated loop profiles. We open source the benchmark data and code, including all tasks, more than 5{,}300 development units, and executable tests, at \href{https://github.com/microsoft/Loopsbench}{\texttt{microsoft/Loopsbench}}.
\end{abstract}

\vspace{-4pt}
\section{Introduction}
\label{sec:introduction}

Current coding agent systems increasingly expose loop mechanisms for sustained software work, e.g., Codex goal mode, Claude Code goal mode, and Claude Code dynamic workflows~\cite{openai_codex_prompting,anthropic_claude_code_goal,anthropic_claude_code_workflows,anthropic_claude_code_subagents}. These mechanisms do not replace the harness. They add a higher level control surface over it, so objectives, progress criteria, and work distribution can persist across extended execution. The core challenge therefore moves from harness engineering alone to loop engineering over the harness. In long-horizon coding, the loop must govern execution across task structure, state continuity, and regression pressure as dependent work accumulates~\cite{anthropic_c_compiler}.

\parabf{Limitations of Existing Benchmarks.} Existing benchmarks reflect this regime only partially. SWE-bench~\cite{jimenez2023swe} and its variants broaden repository level issue resolution along freshness, language coverage, and patch scale~\cite{zan2025multi,yang2024swe,liu2025swebenchm}. Feature level benchmarks~\cite{zhou2026featurebench,chen2025featbench,thai2025swe} further expand the requested change. Their task abstraction nevertheless remains largely terminal: agents receive self contained issues or flat specifications and are judged by final task success. This design measures issue resolution ability but does not reveal whether an agent preserves intermediate obligations, avoids regressions, or follows a viable order through dependent subproblems. Diagnostic long horizon evaluation should expose intermediate development units, track accumulated obligations, and make execution order observable.

\begin{figure}[!t]
  \centering
  \begin{minipage}[c]{0.46\textwidth}
    \centering
    \includegraphics[width=0.85\linewidth]{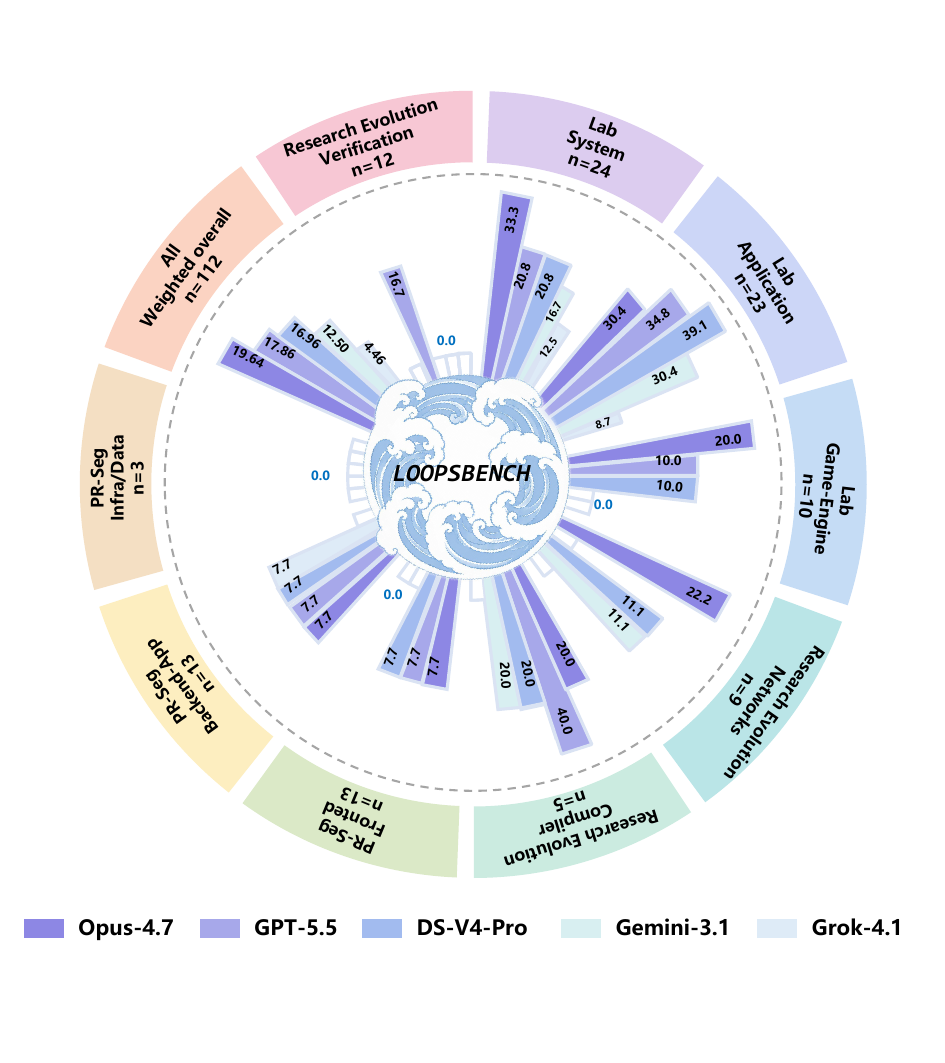}
  \end{minipage}\hfill
  \begin{minipage}[c]{0.50\textwidth}
    \centering
    \includegraphics[width=\linewidth]{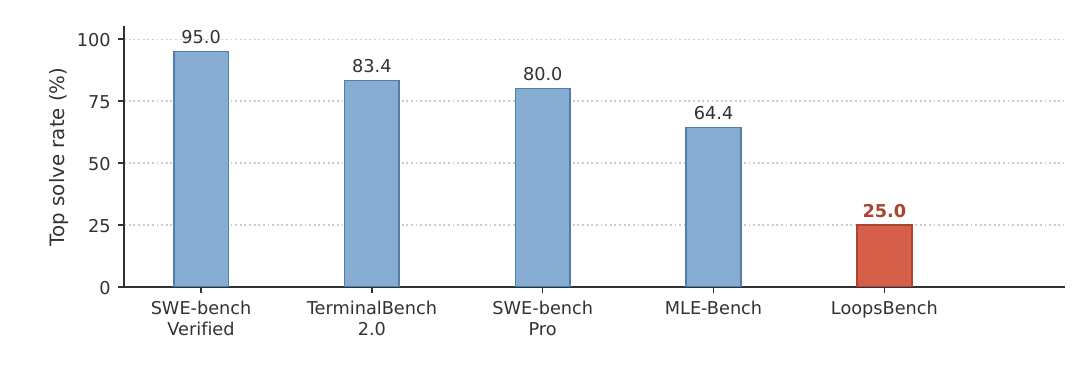}\\[2pt]
    \includegraphics[width=\linewidth]{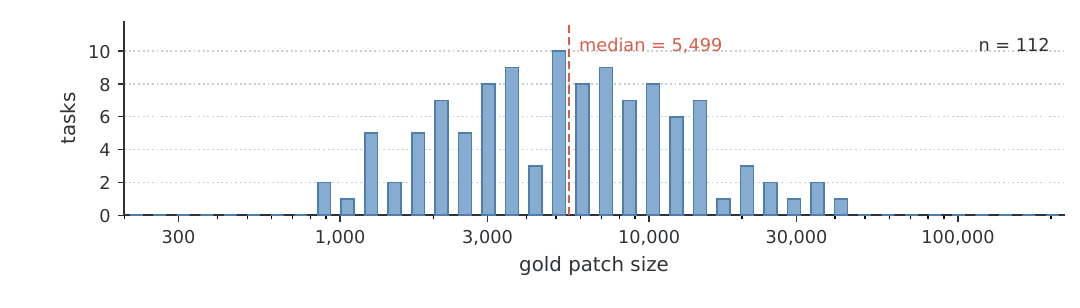}
  \end{minipage}
  \caption{Summary statistics and results for \lhb{}.}
  \label{fig:headline}
\end{figure}

\parabf{\lhb{} as a Benchmark for Loop Engineering Evaluation.} \lhb{} (Figure~\ref{fig:headline}) addresses these requirements by representing each task as a dependency DAG whose nodes are source grounded development units and whose edges encode prerequisite relations. The graph makes intermediate units separately testable and provides a source recovered development order as a descriptive reference for sequencing analysis, not as a claim of optimality. The flow aware evaluation runtime releases ready frontier tests, keeps completed nodes active as regression obligations, and records a loop trace while leaving execution order open.
\lhb{} contains 112 dependency-structured tasks from three authentic, real-world sources, with 29 PR Sequences, 57 Course Labs, and 26 Research Evolutions spanning 9 domains and 8 programming languages. The benchmark has a median dependency depth of 6 and more than 5{,}300 development units. Table~\ref{tab:benchmark-comparison} distinguishes benchmarks that expose separately testable intermediate units from those that only specify terminal tasks, and further separates unit exposure from explicit prerequisite DAGs. This comparison situates \lhb{} as, to our knowledge, the first benchmark for loop engineering evaluation, pairing explicit unit dependency DAGs with a flow-aware harness and loop trace diagnostics.
\providecolor{lhbRuleGray}{HTML}{BDBDBD}
\providecolor{lhbHeaderBlue}{RGB}{232,241,250}
\providecolor{lhbStripe}{RGB}{250,251,253}

\begin{table}[!tb]
  \centering
  \caption{Comparison of \lhb{} with existing coding benchmarks.}
  \label{tab:benchmark-comparison}
  \vspace{3pt}
  \begingroup
  \setlength{\tabcolsep}{3.2pt}
  \setlength{\arrayrulewidth}{0.4pt}
  \renewcommand{\arraystretch}{1.08}
  \footnotesize
  \arrayrulecolor{lhbRuleGray}
  \resizebox{\textwidth}{!}{%
  \begin{tabular}{@{}l l c l c c r r r@{}}
    \toprule
    \rowcolor{lhbHeaderBlue}
    \textbf{Benchmark} &
    \textbf{Evidence} &
    \textbf{Lng} &
    \textbf{Task Form} &
    \textbf{Units} &
    \textbf{DAG} &
    \textbf{Pch} &
    \textbf{Tst} &
    \textbf{Time} \\
    \midrule
    \rowcolor{lhbStripe}\textbf{\lhb{}} & loop trace metrics & 8  & dependency DAG & \benchYes & \benchYes & 37{,}296 & 74  & 6.6m \\
    \midrule
    \rowcolor{lhbHeaderBlue}\multicolumn{9}{@{}l}{\emph{SWE family}} \\
    SWE-bench        & end state F2P & 1  & issue              & \benchNo      & \benchNo & 33      & 9   & 24.6d \\
    \rowcolor{lhbStripe}Multi-SWE-bench  & end state F2P & 7  & issue              & \benchNo      & \benchNo & 107     & 21  & 18.5d \\
    SWE-bench Pro    & end state F2P & 11 & issue or feature & \benchNo      & \benchNo & 462     & 38  & 1.7m \\
    \rowcolor{lhbStripe}FeatureBench     & end state F2P & 1  & feature            & \benchPartial & \benchNo & 1{,}256 & 38  & 19.5d \\
    SWE-EVO          & end state F2P & 1  & software evolution & \benchPartial & \benchNo & 610     & 31  & 7.1d \\
    \midrule
    \rowcolor{lhbHeaderBlue}\multicolumn{9}{@{}l}{\emph{Long horizon family}} \\
    MLE-Bench        & end state F2P & 1  & repo generation    & \benchNo      & \benchNo & 514     & 1   & 3.0m \\
    \rowcolor{lhbStripe}NL2Repo-Bench    & end state F2P & 1  & repo generation    & \benchPartial & \benchNo & 1{,}623 & 247 & 3.4y \\
    ProgramBench     & end state F2P & 1  & repo generation    & \benchPartial & \benchNo & 2{,}104 & 52  & 12.4d \\
    \rowcolor{lhbStripe}RepoZero         & end state F2P & 1  & repo generation    & \benchPartial & \benchNo & 3{,}287 & 68  & 21.7d \\
    \bottomrule
  \end{tabular}%
  }
  \arrayrulecolor{black}
  \endgroup
\end{table}

\parabf{Takeaways.} Three findings emerge. First, the strongest model and loop configuration resolves only 25.00\% of \lhb{}. Second, evaluated loops omit prerequisite relations, produce longer patches, and author sparse tests. Third, context renewal differs across loop implementations, while regression events remain visible in every profile, identifying routing and state tracking as central limitations.

\parabf{Contributions.} \lhb{} provides a graph structured evaluation contract, a scalable task construction pipeline, and 112 source grounded long horizon tasks. Its trace based analysis separates model and loop effects while measuring planning, implementation, testing, routing, and state retention.

\vspace{-4pt}
\section{\lhb{} as a Pipeline for Loop Engineering Evaluation in Long Horizon Coding}\label{sec:design}

\begin{figure}[!t]
  \centering
  \includegraphics[width=\textwidth]{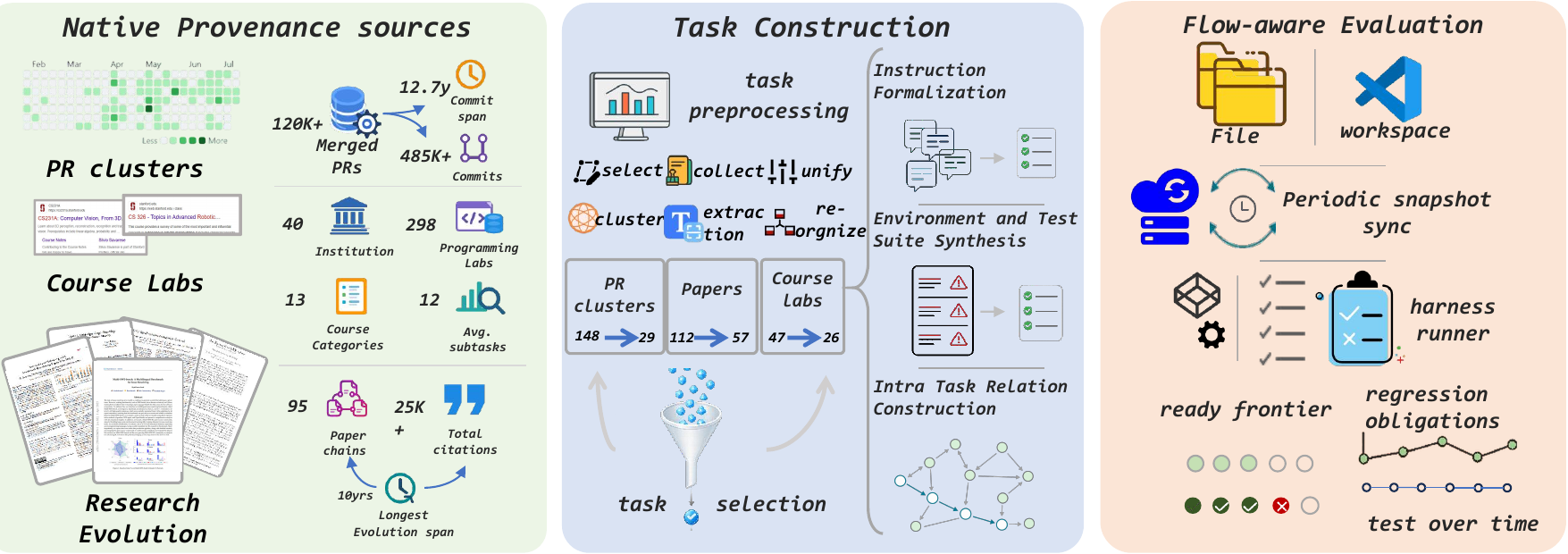}
  \caption{Overview of the \lhb{} construction and evaluation pipeline.}
  \label{fig:overview}
\end{figure}

\rev{Figure~\ref{fig:overview} summarizes the pipeline components, while the numbered stages below specify their construction order.} \lhb{} constructs loop engineering evaluation instances in which dependency structure controls test release, completed work remains live as regression obligations, and each run records loop trace metrics for planning, routing, verification, state continuity, and obligation retention. \textbf{(1) Task Collection} gathers source artifacts from three authentic real world sources. \textbf{(2) Task Preprocessing} normalizes the source artifacts into atomic candidate tasks. \textbf{(3) Task Selection} filters candidates by two source agnostic thresholds.
\rev{\textbf{(4) Development Unit Definition and Intra Task Relation Recovery} defines the units and recovers the dependency DAG that governs ready frontier release.}
\rev{\textbf{(5) Task Instrumentation} then attaches the public instruction, executable environment, and per unit test obligations using the recovered graph.}

\vspace{-3pt}
\subsection{Task Collection}\label{sec:task-collection}

We collect source artifacts from three authentic real world sources spanning a broad spectrum of coding domains, task shapes, and dependency depths. This source mix varies routing burden and obligation retention pressure while keeping all tasks under the same loop engineering evaluation contract. \textbf{(1) Course Labs.} We collect 112 university programming projects from the official course homepages of 30 universities and from public GitHub mirrors of their assignment releases, retrieving each as its complete native bundle of handout, skeleton code, and instructor side reference tests. \textbf{(2) PR Sequences.} We pull the full commit history and merged pull request stream of 56 actively maintained GitHub repositories, all under continuous community development with high star counts and long maintenance windows. \textbf{(3) Research Evolutions.} We gather 212 papers from 48 venues together with the arXiv tex of their cited prior work and forward citers, restricted to papers that are simultaneously highly cited and accompanied by a fully open source official implementation under an OSI approved license. A reference filter retains the 95 papers connected by load bearing inheritance edges.
The three sources jointly anchor the benchmark in authentic developer, educator, and researcher artifacts and yield complementary long horizon structures for evaluating coding agent loops. Per source coverage appears in Appendix~\ref{app:per-task-stats}.

\vspace{-3pt}
\subsection{Task Preprocessing}\label{sec:preprocessing}

We normalize the collected artifacts into atomic candidate tasks before selection so later stages can represent heterogeneous sources through a uniform evaluation contract.

\parabf{Course Labs.} \ding{182}~unify handouts written in mixed languages into a single English version; \ding{183}~reorganize scattered skeleton files under one task root. Each project becomes one candidate task, yielding 112 candidates across the collected projects.

\parabf{PR Sequences.} \ding{182}~clone each repository with full git history and snapshot the first commit as the base codebase; \ding{183}~collect its merged pull requests and extract per pull request diffs split by role (e.g., test, source, auxiliary); \ding{184}~run a Claude Code based extraction step~\cite{claude_code} that performs repository level exploration and emits aggregated segments validated against a fixed acceptance standard. The procedure yields 148 candidate tasks across the 56 repositories.

\parabf{Research Evolutions.} \ding{182}~select a high impact early seed paper in the collected venue set and collect the arXiv tex of the seed together with its cited prior work and forward citers; \ding{183}~run a Claude Code based classification step that labels each reference as a shallow related work mention or a load bearing inheritance edge. \ding{184}~absorb every neighbor connected by a load bearing edge into the cluster and repeat the classification on the newly added papers until the closure admits no further additions. Each closed cluster becomes one candidate task, yielding 47 candidates across the 95 retained papers; 26 survive Selection.
Across all three sources, heterogeneous native test runners are wrapped behind a uniform task entry point, ensuring that every evaluated loop observes the same runtime contract while the source specific dependency evidence is preserved.
Appendix~\ref{app:example-task} presents one materialized \lhb{} task, from candidate recovery to the loop trace metrics recorded during evaluation.

\vspace{-3pt}
\subsection{Task Selection}\label{sec:dedup-quality}

Candidate tasks pass two source agnostic thresholds that retain workloads of sufficient scale for routing burden, state continuity, and regression obligation retention to become observable. \parabf{Temporal Span} is at least 2.5 months, measured as the handout declared project duration for Course Labs, the merge time interval between the first and last commit of the region for PR Sequences, and the publication date span between the earliest and latest paper in the evolution for Research Evolutions. \parabf{Solution Scale} is at least 1{,}200 in the source specific scale. Each source has its own natural patch length and base codebase size. Per source values are reported in Appendix~\ref{app:per-task-stats}.
After filtering, 112 tasks remain as the released loop engineering evaluation suite.

\vspace{-3pt}
\subsection{Intra Task Relation Recovery}\label{sec:dag-framework}

\rev{A materialized development unit is a source grounded, separately testable acceptance unit $u=(r_u,s_u,p_u,\Delta_u,T_u)$, where $r_u$ is its requirement, $s_u$ its file or symbol scope, $p_u$ its prerequisite set, $\Delta_u$ its reference patch contribution, and $T_u$ its standard tests. Unit boundaries follow source provenance: merged PRs or evidence supported PR segments for PR Sequences, released modules or milestones for Course Labs, and load bearing methodological increments for Research Evolutions. Changes that cannot be separated without breaking the acceptance contract are assigned to the nearest prerequisite unit or merged rather than split artificially.}

We formalize each task as a directed acyclic graph over its development units, allowing loop engineering evaluation to observe how a loop routes work through unit level prerequisites. \rev{The fixed DAG is an evaluation contract rather than a claim that real development is always monotonic or interface stable. An evaluated loop may revise earlier implementations provided that the completed predecessor obligations remain satisfied; tasks that intentionally replace earlier requirements through exploratory redesign fall outside this abstraction.} The four admitted patterns are \emph{sequential PR chains} along the merged commit history, \emph{structural module reuse} when $v$ edits a file or symbol that $u$ introduced, \emph{functional producer consumer API edges} when $v$ calls or imports an authoritative definition created by $u$, and \emph{compositional layering} when $v$ extends a subclass, schema, or interface declared by $u$.

\begin{center}
  \begin{minipage}{\textwidth}
    \centering
    \includegraphics[width=\linewidth]{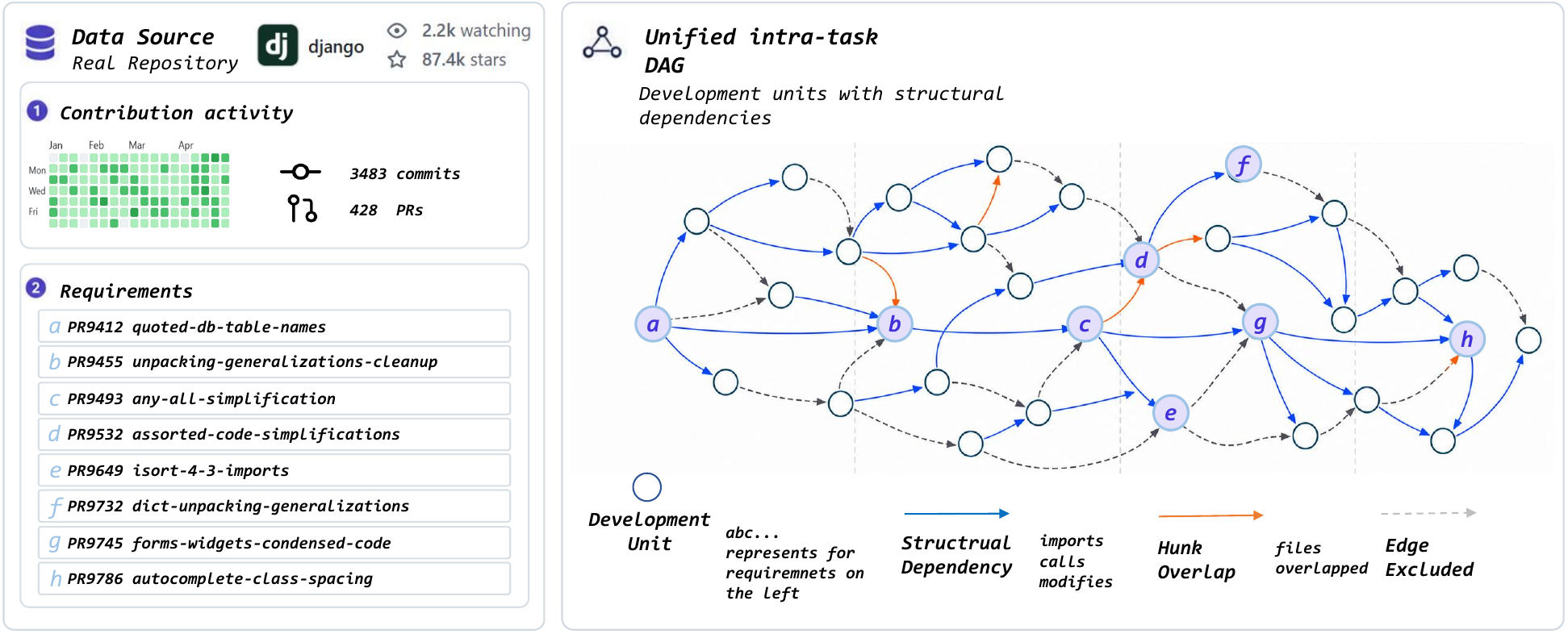}
    \captionof{figure}{Four prerequisite patterns in a materialized task.}
    \label{fig:dependency-types}
  \end{minipage}
\end{center}

An edge is admitted only under unambiguous evidence on the gold artifacts. The first kind of evidence is hunk overlap on the same file. The second kind of evidence is symbol level dependence through an import, call, or subclass of an authoritative definition introduced by $u$. Spurious chains from hot monoliths, generated artifacts, and lockfiles are excluded by a denylist applied before edge admission. A file touched by most PRs in a repository therefore contributes no edges, and a regenerated lockfile never anchors a successor. The full admission rules appear in Appendix~\ref{app:task-schema}. On $G$, the \emph{ready frontier} at checkpoint $t$ is
\[
R_t \;=\; \{\, v \in V \setminus C_t \;:\; \forall (u,v) \in E,\; u \in C_t \,\},
\]
where $C_t$ collects nodes whose tests all pass at $t$.
For loop engineering evaluation, edges encode how local requirements become persistent obligations that remain active as later units are attempted.
The tests for each unit are evaluated in a workspace where predecessor tests remain enforced, measuring progress on later units together with preservation of earlier obligations.

\vspace{-3pt}
\subsection{Task Instrumentation}\label{sec:description-refine}

We turn each selected candidate into a self contained evaluation instance whose evaluation contract contains a formalized task instruction, a reproducible environment, and discriminative tests. The gold solution stays outside the evaluated loop context and is used only to validate executable obligations and compute reference statistics before evaluation begins.

\ding{182}~\textbf{Instruction Formalization.} We treat the gold diff against the base codebase as ground truth and recover the developer's acceptance intent from it, since PR titles, course handouts, and paper method sections rarely specify the full executable contract on their own.
A Claude Code based recovery step~\cite{claude_code} backed by Claude Opus 4.7~\cite{anthropic_opus47} produces a single acceptance instruction while keeping the implementation path hidden.
The recovery step has two roles in the evaluation contract. It removes evidence of the implementation path, including concrete edit locations and gold diff structure. It retains the public acceptance contract that the test suite binds to, including public signatures and expected output formats.
A symbol enters the instruction when a test assertion depends on naming it.
The same recovery step strips repository names, course identifiers, paper titles, and author attributions to reduce prompt level source recognition.

\ding{183}~\textbf{Environment and Test Suite Synthesis.}
A Claude Code based environment synthesis step traverses the source frame from upstream preprocessing, collects runtime dependencies, and writes a \texttt{Dockerfile} for the environment, with a companion \texttt{docker-compose.yaml} when the task spans multiple services.
\rev{After development units and their dependency DAG have been recovered,} the materialization step traverses the DAG in topological order so each unit test is authored against the prerequisite state used for ready frontier release.
At each unit $u$, the materialization pipeline drafts missing tests on top of the current environment, applies $u$'s gold patch on the recorded base, and runs the new tests under a three trial fail to pass check that admits $T(u)$ when the full gold solution flips every test to pass (\emph{solvability}), the empty workspace flips none (\emph{non triviality}), and the gold solution restricted to $u$'s DAG ancestors still leaves at least one test failing (\emph{discriminativeness} with respect to $u$'s own contribution).
A test with no execution result is attributed to the environment and triggers a \texttt{Dockerfile} revision, while a test that executes yet remains nonpassing under the gold patch is attributed to the assertion and rewritten in place.
After every unit has been visited, a closure pass replays the full gold patch against the accumulated suite on the recorded base. This grounds loop traces in executable obligations and excludes textual milestones from the scoring contract.
Units with unrepaired environment or suite issues after the computational budget are discarded.

\vspace{-3pt}
\subsection{Evaluation Protocol}\label{sec:protocol}

\begin{figure}[!t]
  \centering
  \includegraphics[width=0.85\linewidth]{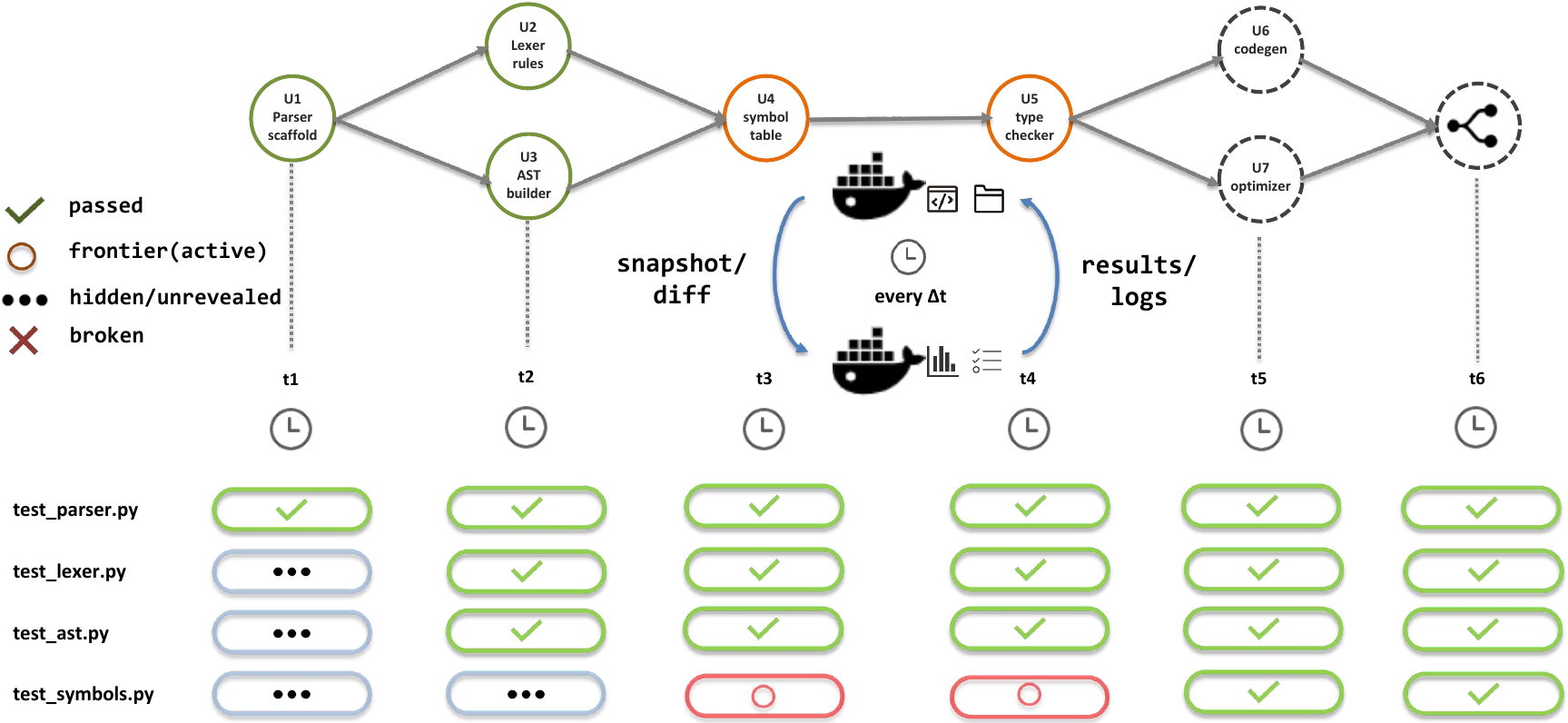}
  \caption{Flow aware evaluation runtime.}
  \label{fig:protocol}
\end{figure}

As illustrated in Figure~\ref{fig:protocol}, the evaluated loop receives the task instruction and per unit requirement bundles as plain files, while gold solutions and the ready unit identity remain withheld. \rev{Checkpoint outcomes and the active obligation state are retained for evaluator side scoring; the evaluated loop can inspect the attached tests, use project native tests, and author tests of its own. Appendix~\ref{app:task-schema} gives the complete visibility contract in Table~\ref{tab:visibility-contract}.} \rev{The evaluation runtime records the visitation sequence in the loop trace and compares the prerequisite structure in the recorded plan with the source recovered DAG. It accepts any valid topological execution order rather than requiring reproduction of the historical total order.}

\parabf{Topologically Gated Test Release.}
The evaluation runtime releases tests layer by layer along the dependency DAG, with every multi predecessor node acting as a gate.
\rev{Here, release and sealing refer only to evaluator side activation for scoring, not disclosure of the attached test files or bindings.}
A gate node $v$ joins the ready frontier after all predecessors lie in $C_t$, and the tests of its descendants stay sealed until $T(v)$ flips from fail to pass.
The gate affects scoring rather than editing permission. The evaluated loop remains free to edit across layers, and the recorded loop trace reports whether its work satisfies the dependency frontier without being told which unit is currently ready.
Once a unit clears the gate, its tests and the tests of its predecessors are kept enforced as regression tests on every later layer, scoring subsequent edits against both new work and completed obligations.
\rev{Regression Rate therefore measures preservation of previously satisfied behavior without staged evaluator feedback about checkpoint outcomes or the active obligation state.}

\parabf{Dual Container Snapshot Pipeline.}
We separate edit execution from test adjudication across two containers, allowing the evaluated loop to continue editing while the evaluation runtime independently samples workspace state and updates released obligations.
Container A hosts the working tree and is the sole write target for the evaluated loop.
Container B contains the full test suite and the reference environment, and reads from A through snapshots managed by the evaluation runtime.
On a fixed time cadence, the evaluation runtime samples A and diffs the new snapshot against the queue head. Empty diffs are dropped, while non empty diffs are appended to the queue and shipped into B for the currently released layer to run against.
Snapshots form a queue of monotonic edit states, allowing \lhb{} to compute loop trace metrics at real change points as well as at the final state.

\vspace{-4pt}
\section{Experiments}\label{sec:experiments}

\definecolor{macaronPinkHead}{HTML}{F4B6C2}
\definecolor{macaronPinkBody}{HTML}{FCE4EC}
\definecolor{macaronBlueHead}{HTML}{B6D4F4}
\definecolor{macaronBlueBody}{HTML}{E3F0FC}
\definecolor{macaronMintHead}{HTML}{B7E4C7}
\definecolor{macaronMintBody}{HTML}{E6F5EC}
\definecolor{macaronLemonHead}{HTML}{F7E29A}
\definecolor{macaronLemonBody}{HTML}{FBF3D0}
\definecolor{macaronLavHead}{HTML}{D4C2F0}
\definecolor{macaronLavBody}{HTML}{EEE7F8}

\definecolor{lhbAnswerBg}{RGB}{243,248,252}
\newenvironment{summary}{%
  \begin{tcolorbox}[width=\linewidth, colback=lhbAnswerBg, top=1pt, bottom=1pt, left=2pt, right=2pt, before skip=3pt, after skip=3pt]%
}{%
  \end{tcolorbox}%
}

We organize the experiments around three questions that treat coding agent evaluation as loop engineering evaluation.
All experiments use the full 112 task release of \lhb{}.
\vspace{-3pt}
\subsection{RQ1: How Far Do Coding Agent Loops Sustain Progress?}\label{sec:rq1}

\autoref{tab:rq1} reports RQ1 results along three dimensions that separate model choice, loop implementation, and residual continuation. The table covers within family model scaling under the corresponding vendor loop, model sweeps under a fixed loop implementation, and loop implementation sweeps under a fixed model.
Across all three dimensions, every cell remains well below full resolution of \lhb{}, suggesting that long horizon limitations are not explained by model choice or loop implementation in isolation.
The strongest model and loop configuration resolves 25.00\% of tasks. Without the outer evaluation loop, a single evaluated execution segment usually reaches a shallow dependency prefix.
These pairings align frontier models with the vendor infrastructure under which they are typically deployed, yet long horizon loop execution remains well below complete resolution.
The loop implementation sweep isolates how the surrounding execution loop changes progress under the same model.
Without external continuation, evaluated loop implementations cover a shallow prefix under their own state retention and work routing policies.
The recorded loop traces also capture partial progress patterns, including unresolved units with empty patches and skipped obligations that prioritize lower burden units.
External continuation reduces early stalls and increases partial progress, while residual routing and state retention remain central loop engineering factors.
In the model dimension, Resolve Rate plateaus even when the outer evaluation loop attempts every unit, because later units inherit earlier open obligations.
Even under tool use and execution oriented training, the evaluated model and loop configurations often decompose the task into local issues without consistently maintaining a persistent global plan. Appendix~\ref{app:passk} reports pass@K to check that within family monotonicity is not driven by a single trial. In Table~\ref{tab:rq1}, RR denotes Resolve Rate, TPR denotes Test Pass Rate, and Depth is normalized dependency depth. Mean billed tokens per task appear in Appendix~\ref{app:rq1-tokens}.

\begin{table*}[!t]
  \centering
  \caption{RQ1 performance across models and loop implementations. Per task billed tokens are reported in Appendix~\ref{app:rq1-tokens}.}
  \label{tab:rq1}
  \begin{minipage}[t]{0.49\textwidth}
    \vspace{0pt}
    \centering
    \footnotesize
    \setlength{\tabcolsep}{5pt}
    \renewcommand{\arraystretch}{1.16}
    \resizebox{\linewidth}{!}{%
    \begin{tabular}{@{}l cc cc c@{}}
      \toprule
      \textbf{Model} & \multicolumn{2}{c}{\textbf{RR}$\uparrow$} & \multicolumn{2}{c}{\textbf{TPR}$\uparrow$} & \textbf{Depth}$\uparrow$ \\
       & \textit{w/o} & \textit{w/} & \textit{w/o} & \textit{w/} & \\
      \midrule
      \rowcolor{macaronMintHead}\multicolumn{6}{l}{\emph{GPT (Codex)}} \\
      GPT-5.5       & 14.29\% & 21.43\% & 39.62\% & 50.84\% & 0.53 \\
      GPT-5.2       & 9.82\%  & 13.39\% & 33.91\% & 43.90\% & 0.36 \\
      GPT-5         & 5.36\%  & 8.04\%  & 28.45\% & 36.86\% & 0.22 \\
      GPT-4o        & 1.79\%  & 3.57\%  & 21.85\% & 28.30\% & 0.09 \\
      \rowcolor{macaronMintHead}\multicolumn{6}{l}{\emph{Claude (Claude Code)}} \\
      Opus-4.7      & 16.96\% & 25.00\% & 41.18\% & 53.05\% & 0.61 \\
      Opus-4.5      & 12.50\% & 17.86\% & 36.97\% & 48.21\% & 0.46 \\
      Sonnet-4.5    & 8.93\%  & 12.50\% & 32.85\% & 42.58\% & 0.34 \\
      Sonnet-4      & 4.46\%  & 7.14\%  & 27.43\% & 35.54\% & 0.18 \\
      \rowcolor{macaronMintHead}\multicolumn{6}{l}{\emph{Qwen (Qwen Code)}} \\
      Qwen3.6-Plus  & 6.25\%  & 9.82\%  & 37.97\% & 45.10\% & 0.31 \\
      Qwen3.5-Plus  & 5.36\%  & 8.04\%  & 31.75\% & 41.17\% & 0.19 \\
      Qwen3-Max     & 0.89\%  & 1.79\%  & 17.45\% & 23.31\% & 0.07 \\
      Qwen2.5-72B   & 0.00\%  & 0.00\%  & 10.61\% & 13.88\% & 0.03 \\
      \noalign{\vskip 6.4pt}
      \bottomrule
    \end{tabular}%
    }
  \end{minipage}\hfill
  \begin{minipage}[t]{0.49\textwidth}
    \vspace{0pt}
    \centering
    \footnotesize
    \setlength{\tabcolsep}{5pt}
    \renewcommand{\arraystretch}{1.16}
    \resizebox{\linewidth}{!}{%
    \begin{tabular}{@{}l cc cc c@{}}
      \toprule
      & \multicolumn{2}{c}{\textbf{RR}$\uparrow$} & \multicolumn{2}{c}{\textbf{TPR}$\uparrow$} & \textbf{Depth}$\uparrow$ \\
       & \textit{w/o} & \textit{w/} & \textit{w/o} & \textit{w/} & \\
      \midrule
      \rowcolor{macaronPinkHead}\multicolumn{6}{l}{\textbf{Model} \textit{(fixed loop: Claude Code)}} \\
      Opus-4.7      & 16.96\% & 25.00\% & 41.18\% & 53.05\% & 0.61 \\
      GPT-5.5       & 13.39\% & 20.54\% & 38.04\% & 49.30\% & 0.47 \\
      GLM-5.1       & 13.39\% & 18.75\% & 34.55\% & 45.49\% & 0.44 \\
      DeepSeek-V4P  & 11.61\% & 18.75\% & 36.07\% & 47.34\% & 0.46 \\
      Gemini-3.1-Pro & 8.93\%  & 14.29\% & 32.86\% & 43.59\% & 0.31 \\
      Qwen3.6-Plus  & 6.25\%  & 9.82\%  & 36.43\% & 47.50\% & 0.34 \\
      Kimi-2.6      & 3.57\%  & 6.25\%  & 21.27\% & 28.33\% & 0.14 \\
      Grok-4.1-FR   & 2.68\%  & 4.46\%  & 23.32\% & 30.97\% & 0.15 \\
      \rowcolor{macaronBlueHead}\multicolumn{6}{l}{\textbf{Loop} \textit{(fixed model: \texttt{gpt-5.4})}} \\
      Codex          & 13.39\% & 18.75\% & 37.84\% & 49.06\% & 0.49 \\
      Claude Code    & 12.50\% & 17.86\% & 36.97\% & 48.21\% & 0.42 \\
      GitHub Copilot & 10.71\% & 15.18\% & 34.21\% & 45.46\% & 0.38 \\
      OpenHands      & 6.25\%  & 9.82\%  & 29.21\% & 39.30\% & 0.20 \\
      SWE-agent      & 5.36\%  & 8.93\%  & 28.07\% & 38.11\% & 0.16 \\
      mini-swe-agent & 4.46\%  & 7.14\%  & 25.96\% & 35.20\% & 0.13 \\
      \bottomrule
    \end{tabular}%
    }
  \end{minipage}
\end{table*}

\begin{summary}
\textbf{RQ1:} Sustained long horizon loop progress remains limited at frontier scale. The limitation appears in loop level obligations such as global plan maintenance, residual continuity, and regression obligation retention rather than in per unit code generation alone.
\end{summary}

\vspace{-3pt}
\subsection{RQ2: How Do Coding Agent Loops Maintain Plan, Code, and Test State?}\label{sec:rq2}
\begin{table}[t]
  \centering
  \caption{Planning, implementation, and testing loop trace metrics.}
  \label{tab:rq2-grid}
  \definecolor{rqGroupRule}{HTML}{ECECEC}
  \tiny
  \setlength{\tabcolsep}{3.6pt}
  \renewcommand{\arraystretch}{1.05}
  \resizebox{\linewidth}{!}{%
  \begin{tabular}{@{}l cccc cc ccc @{}}
    \toprule
    & \multicolumn{4}{c}{\textbf{Planning fidelity}}
    & \multicolumn{2}{c}{\textbf{Implementation}}
    & \multicolumn{3}{c}{\textbf{Testing}} \\
    \cmidrule(lr){2-5} \cmidrule(lr){6-7} \cmidrule(lr){8-10}
    \textbf{Loop}
      & \textbf{Edge F1} & \textbf{Layer~$\rho$} & \textbf{CPR} & \textbf{WR}
      & \textbf{PatchLen} & \textbf{Jacc}
      & \textbf{\#T} & \textbf{F2P} & \textbf{Reg}$\downarrow$ \\
    \midrule
    \rowcolor{rqGroupRule}\multicolumn{10}{l}{\emph{Closed source loop implementations}} \\
    Claude Code
      & 0.71 & 0.65 & 0.31 & 1.08
      & 1.58 & 0.62
      & 28 & 0.47 & 7.11\% \\
    Codex
      & 0.67 & 0.61 & 0.33 & 1.14
      & 1.71 & 0.64
      & 24 & 0.44 & 4.83\% \\
    GitHub Copilot
      & 0.58 & 0.52 & 0.41 & 0.92
      & 1.83 & 0.66
      & 22 & 0.41 & 6.91\% \\
    \rowcolor{rqGroupRule}\multicolumn{10}{l}{\emph{Open source loop implementations}} \\
    OpenHands
      & 0.39 & 0.33 & 0.85 & 0.39
      & 2.31 & 0.73
      & 16 & 0.34 & 2.46\% \\
    SWE-agent
      & 0.37 & 0.31 & 0.88 & 0.36
      & 2.36 & 0.74
      & 15 & 0.33 & 2.18\% \\
    mini-swe-agent
      & 0.27 & 0.22 & 0.97 & 0.24
      & 2.54 & 0.76
      & 11 & 0.28 & 0.24\% \\
    \bottomrule
  \end{tabular}%
  }
\end{table}

\rev{\autoref{tab:rq2-grid} uses recorded loop traces to compare evaluated loops with the source recovered prerequisite DAG along planning, implementation, and testing axes. The runtime permits any topological execution order consistent with the DAG, while the planning diagnostics score source evidenced relations and layers rather than reproduction of a historical total order.}
\textbf{(1) Planning: an internal dependency gap.}
\rev{Recorded plans recover partial dependency structure for long horizon routing, omitting many source evidenced pull request level prerequisites in the reference DAG.}
\rev{Edge F1 and Layer~$\rho$ remain far from perfect agreement with the source recovered prerequisite DAG for every evaluated loop, with lower values for open loop implementations than for closed ones.}
Critical Path Ratio (CPR) and Width Ratio (WR) reveal how missing prerequisites reshape the planned concurrency budget.
\rev{Closed source loop implementations organize execution through a tree shaped concurrent structure and stay close to the concurrency budget implied by the reference DAG, whereas open source loop implementations driven by linear control flow form a near chain.}
\rev{Width Ratio above unity for closed loop implementations is better interpreted as overparallelization than as planning fidelity beyond the source recovered reference DAG.}
A limited share of those branches reflects valid alternative decomposition, while most parallelize serial requirements absent from the recorded plan.
\textbf{(2) Implementation: patch surplus over a comparable token base.}
On units that the evaluated configuration eventually resolves, PatchLen exceeds the gold reference by a consistent margin while token overlap with gold remains moderate. Candidate patches are therefore often longer than the reference while preserving enough token overlap to remain comparable to it. Appendix~\ref{app:rq2-extended} provides a diff level account of where the extra lines come from.
\textbf{(3) Testing: a maintenance gap over the long horizon.}
Recorded loop traces contain sparse test authoring as long horizon work unfolds, leaving earlier obligations with limited agent authored regression protection against later edits.
Test count (\#T) is markedly lower than that of the native suite for every evaluated loop. Fail to pass yield (F2P) on authored tests remains acceptable after internal validation before commit. Regression (Reg) captures the downstream cost of this sparse maintenance.
Deployed loop implementations exhibit several percentage points of regression over previously satisfied obligations. The lower apparent Regression Rate of lighter implementations reflects the limited number of passing obligations eligible for regression measurement.
Long horizon loop engineering evaluation derives loop trace metrics from the running state of the evaluated loop, recording where plan, code, and test state are maintained, rerouted, or degraded.
\begin{summary}
\textbf{RQ2:} The long horizon gap centers on loop state discipline. Recorded loop traces capture missing dependencies, inflated patches, and sparse tests, linking performance to how plan, code, and test state are maintained across the run.
\end{summary}

\vspace{-3pt}
\subsection{RQ3: Which Loop Engineering Factors Shape Long Horizon Execution?}\label{sec:rq3}

RQ3 studies objective persistence, residual routing, context renewal, and regression pressure. We compare four representative loop runs. \emph{Codex goal mode} keeps a persistent objective inside the Codex loop~\citep{openai_codex_prompting}. \emph{Claude Code goal mode} maintains a session condition checked after each turn~\citep{anthropic_claude_code_goal}. \emph{Claude dynamic workflows} launch task specific workers with narrower local contexts~\citep{anthropic_claude_code_workflows,anthropic_claude_code_subagents}. The Ralph loop starts a fresh invocation for residual work~\citep{huntley_ralph_wiggum,huntley_ralph_loop}.
\autoref{tab:rq3-loop-engineering} reports context-budget rounds, regression events per run, and Resolve Rate. A context-budget round is the initial context or a segment opened by the implementation compaction policy or by an external restart. \autoref{fig:rq3-compaction-ablation} shows the compaction-budget sensitivity. Appendix~\ref{app:loop-compaction-metrics} gives the definitions and run coverage.

\begin{table*}[t]
  \centering
  \begin{minipage}[t]{0.58\textwidth}
    \vspace{0pt}
    \centering
    \captionsetup{skip=5pt}
    \caption{Loop trace metrics.}
    \label{tab:rq3-loop-engineering}
    \footnotesize
    \setlength{\tabcolsep}{4pt}
    \renewcommand{\arraystretch}{1.18}
    \begin{tabular*}{\linewidth}{@{\extracolsep{\fill}}l c c c@{}}
      \toprule
      \textbf{Loop Implementation} & \textbf{Rounds} & \textbf{Reg/run} & \textbf{RR (\%)} \\
      \midrule
      Codex goal mode & 32.76 & 0.34 & 20.59 \\
      Claude Code goal mode & 34.69 & 0.13 & 17.65 \\
      Claude dynamic workflows & 97.96 & 0.36 & \textbf{24.11} \\
      Ralph loop & 13.24 & 0.17 & 7.84 \\
      \bottomrule
    \end{tabular*}
  \end{minipage}\hfill
  \begin{minipage}[t]{0.38\textwidth}
    \vspace{0pt}
    \centering
    \captionsetup{skip=5pt}
    \captionof{figure}{Compaction sensitivity.}
    \label{fig:rq3-compaction-ablation}
    \includegraphics[width=\linewidth]{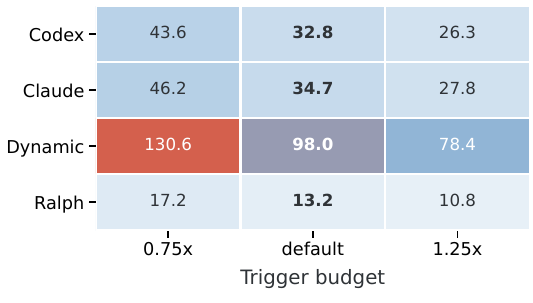}
  \end{minipage}
\end{table*}

\textbf{(1) Context renewal differs across loop implementations.}
After counting each compaction interval as a round, dynamic workflows record 97.96 context-budget rounds per run, compared with 34.69 for Claude Code goal mode, 32.76 for Codex goal mode, and 13.24 for Ralph. The dynamic workflow run has the highest Resolve Rate at 24.11\%, while Ralph reaches 7.84\%. These values describe the evaluated configurations.
\Needspace{4\baselineskip}
\textbf{(2) Context-budget renewal does not remove regression pressure.}
Dynamic workflows record 0.36 regression events per run despite distributing work across narrower worker contexts. Codex goal mode records 0.34, Claude Code goal mode 0.13, and Ralph 0.17. Dynamic routing therefore reduces dependence on a single growing transcript, but completed obligations still require explicit state tracking when workers return. Appendix~\ref{app:rq3-trace-scope} states the trace scope for worker local contexts.
\begin{summary}
\textbf{RQ3:} Loop engineering changes how context is renewed and how residual work is routed. Dynamic workflows achieve the strongest observed outcome, while regression events remain visible across all four loop profiles.
\end{summary}

\vspace{-4pt}
\section{Related Work}\label{sec:related}

Prior coding benchmarks establish reliable end state scoring for code changes, but they offer limited evidence about the loops that sustain long horizon execution. Function level code generation benchmarks~\citep{chen2021evaluating} pioneered programmatic evaluation on localized snippets, while repository level benchmarks centered on SWE-bench~\citep{jimenez2023swe} package work as issue resolution and emphasize end state pass rate.
Feature and evolution benchmarks~\citep{zhou2026featurebench} enlarge task scope, and broader long horizon benchmarks~\citep{kwa2025measuring} extend duration, repository scale, or interaction depth.
Across these families, prerequisite state, residual work routing, and regression obligations generally remain outside the executable scoring contract. Success is therefore assigned to submitted artifacts rather than to the loop state that produced them.
Modern coding agent infrastructure~\citep{claude_code,openai_codex_prompting,anthropic_claude_code_workflows,huntley_ralph_loop} increasingly relies on persistent objectives, worker orchestration, restarts, and context management. Loop implementation therefore becomes part of the evaluated behavior, while current benchmark designs provide limited evidence about how evaluated loops preserve state and route residual work.
Software engineering work on change impact and release planning~\citep{lehnert2011taxonomy} treats dependency structure as an explicit artifact. \lhb{} draws on this view by converting dependency structure into executable obligations for coding agent loop evaluation. Appendix~\ref{app:related-bench} provides the full grouping.

\vspace{-4pt}
\section{Conclusion}\label{sec:conclusion}

\lhb{} makes loop engineering a measurable dimension of coding agent evaluation alongside end state scoring.
It formalizes long horizon tasks as dependency DAGs, releases tests along graph edges, and retains completed unit tests as persistent regression obligations while leaving execution order to the evaluated loop.
This design records loop traces that capture residual work routing, state continuity, and obligation retention across iterations.
Across frontier LLMs and loop implementations, the strongest configuration resolves 25.00\% of tasks, regression events remain visible across all four RQ3 loop profiles, and \rev{recorded plans recover only part of the source recovered prerequisite DAG; the diagnostic does not require reproduction of a particular historical total order.}
These results support evaluating long horizon coding agents with joint attribution to model choice and loop configuration, especially residual work handling, structured state, and regression obligation retention. Scope conditions, source assumptions, and version pinning caveats are stated in the Limitations section.

\section*{Limitations}

The recovered DAG is a lower bound on prerequisite structure rather than a complete causal graph of development. Edge admission uses four source evidenced patterns and a hot file denylist. The resulting depth and width statistics are therefore conservative. \rev{It may miss implicit dependencies in configuration, data formats, build systems, or cross service behavior. The fixed DAG is an evaluation contract: unit boundaries and reference patches are nonunique, and exploratory redesigns that replace earlier requirements fall outside its scope.}
The three pipelines admit tasks when commit history, curriculum decomposition, or citation closure yields verifiable evidence, narrowing the current pool to sources with auditable prerequisite evidence and leaving mobile, frontend heavy, and hardware adjacent projects outside scope.
Reported correctness metrics are tied to the released test suite, including TPR, Resolve Rate, and Regression Rate, and therefore measure executable obligations rather than full semantic equivalence.
\rev{Hidden checkpoint state can make regressions reflect limited feedback as well as regression discipline. Claude Code is used only offline; evaluated loops share one contract and receive no construction transcript, gold solution, or active frontier. Model assisted materialization may still bias wording, boundaries, and tests, while public sources leave contamination risk.}
\lhb{} records coding agent loop execution through traces made accessible by current loop implementations and evaluation adapters. Internal events outside these interfaces cannot be measured directly.
As goal modes, dynamic workflows, compaction policies, and external loop implementations evolve, \lhb{} provides a loop engineering evaluation interface with replaceable adapters and version pinned run metadata.
\rev{Evaluation is costly, and proprietary outputs may drift. We will release tasks, tests, pinned environments, construction and evaluation code, prompts, configurations, traces, checksums, and a smoke test subset to reproduce the contract, though not necessarily exact outputs.}
\rev{\lhb{} measures one executable contract, not general software engineering ability or deployment readiness; generated code still requires human review, security testing, and licensing checks.}

\section*{Compliance, Safety, and Data Statement}

\rev{\lhb{} is constructed from public open source repositories, public course artifacts, and publicly released research code. It does not use personal data, customer data, or private production data. Generated benchmark instances are derived from source artifacts and should preserve the applicable source licenses, attribution requirements, and redistribution constraints. The construction pipeline may reintroduce historical defects, including security relevant defects, as part of benchmark generation. These instances are intended for research evaluation rather than deployment. We assume source histories and release artifacts are benign and trusted. Malicious, poisoned, or intentionally deceptive repositories are outside the scope of this benchmark.}

\bibliographystyle{plainnat}
\bibliography{references}

\appendix

\section{Related Work for Loop Engineering Evaluation}\label{app:related-bench}

Table~\ref{tab:related-bench} groups the full set of prior work surveyed in Section~\ref{sec:related} by the evaluation evidence each cluster supports and by its relation to loop engineering evaluation. Section~\ref{sec:related} cites the lead exemplar of each cluster, while this appendix records additional references without expanding main text citation density.

\begin{table}[!ht]
  \centering
  \caption{Prior work related to loop engineering evaluation.}
  \label{tab:related-bench}
  \tiny
  \setlength{\tabcolsep}{2pt}
  \renewcommand{\arraystretch}{0.74}
  \begin{tabular}{@{}p{0.30\linewidth} p{0.62\linewidth}@{}}
    \toprule
    \textbf{Cluster} & \textbf{References (representative + extensions)} \\
    \midrule
    \textbf{Function level code generation.}
    Report localized unit test success, without state retention or residual routing over long horizons.
      & \citet{chen2021evaluating,hendrycks2021measuring,jain2024livecodebench,liu2023your} \\
    \midrule
    \textbf{SWE-bench family.}

    Repository level issue tasks that primarily report end state success for a submitted patch.
      & \citet{jimenez2023swe,swebench,zan2025multi,yang2024swe,liu2025swebenchm,swebench_pro,swebench_verified,swelancer,swegym,rebench,sonwane2026omnicode,sehgal2025formulacode,tan2024devbench} \\
    \midrule
    \textbf{SWE-bench audits.}

    Document contamination and overfitting risks in verified issue resolution variants.
      & \citet{sbv_contamination,zhu2025establishing} \\
    \midrule

    \textbf{Feature and evolution.}
    Broaden task scope while generally leaving prerequisite structure outside executable obligations that measure evaluated loops.
      & \citet{zhou2026featurebench,chen2025featbench,thai2025swe,gautam2025refactorbench,orlanski2026slopcodebench,longcli_bench,nl2repo_bench} \\
    \midrule

    \textbf{Interaction and trace benchmarks.}
    Extend beyond one submitted change while generally leaving completed units outside active regression obligations during the recorded loop trace.
      & \citet{kwa2025measuring,siegel2024core,li2024devbench,merrill2026terminal,edwards2025rexbench,chen2026beyondswe,vijayvargiya2025interactive} \\
    \midrule

    \textbf{Loop implementations and infrastructure.}
    Systems and frameworks that instantiate coding agent loops through prompts, tools, memory, execution policies, and adapters, forming the system boundary studied by loop engineering evaluation.
      & \citet{claude_code,codex,devin,wang2024openhands,yang2024sweagent,swe_agent,autocoderover,agentless,copilot_cloud_agent,huntley_ralph_wiggum} \\
    \midrule

    \textbf{Loop engineering infrastructure.}
    Describe infrastructure patterns for persistent goals, external loops, dynamic worker orchestration, and subagent routing, but do not define a benchmark over dependency structured obligations.
      & \citet{openai_codex_prompting,anthropic_claude_code_goal,anthropic_claude_code_workflows,anthropic_claude_code_subagents,huntley_ralph_loop} \\
    \midrule
    \textbf{Reasoning frameworks.}
    Provide planning, memory, reflection, and context management primitives that loop implementations build on, rather than executable loop trace evidence.
      & \citet{yao2023tree,besta2024graph,shinn2023reflexion,hong2023metagpt,wu2024autogen,wang2023voyager,packer2023memgpt} \\
    \midrule
    \textbf{Software engineering prerequisites.}
    Study dependency structure, PR ordering, and release planning, providing source concepts that \lhb{} converts into executable obligations and loop trace metrics.
      & \citet{lehnert2011taxonomy,tu2002evolution,costa2016framework,valdivia2018characterizing,ruhe2005art} \\
    \bottomrule
  \end{tabular}
\end{table}
\FloatBarrier
\section{Task Pool Statistics for Loop Engineering Evaluation}\label{app:per-task-stats}

The 112 \lhb{} tasks decompose into 29 PR Sequences (Infra and Data, Backend, Frontend), 57 Course Labs (Game Engines, Systems, Apps and ML), and 26 Research Evolutions (Analysis and Verification, Systems and Networking, Compiler with Database and ML). This source mix spans 26 leaf subcategories and covers software work with different dependency structures, routing burdens, and verification obligations.

Tables~\ref{tab:lhb_task_structure_stats} and~\ref{tab:lhb_task_structure_stats_part2} report per task structural statistics for all 112 \lhb{} tasks, including repository scale (\#Files, \#LoC), unit count (\#Num development units), task structure summaries (average \#Modules, average DAG edges, average dependency depth), and gold implementation targets (average \#Lines, \#Files, \#LoC). These statistics describe the work graph and implementation scope used to quantify routing burden and obligation retention pressure in loop engineering evaluation. The split into two tables is for typesetting, and the row sets are disjoint and cover the full task pool.

\begin{table}[!htbp]
\centering
\caption{Structural statistics for \lhb{} tasks, Part 1.}
\label{tab:lhb_task_structure_stats}
\scriptsize
\setlength{\tabcolsep}{3.0pt}
\renewcommand{\arraystretch}{1.05}
\resizebox{\textwidth}{!}{%
\begin{tabular}{@{}lrr|r|rrr|rrr@{}}
\toprule
\multicolumn{3}{c|}{\textbf{Repository}} &
\multicolumn{1}{c|}{\textbf{Units}} &
\multicolumn{3}{c|}{\textbf{Task structure}} &
\multicolumn{3}{c}{\textbf{Implementation target}} \\
\cmidrule(lr){1-3}
\cmidrule(lr){4-4}
\cmidrule(lr){5-7}
\cmidrule(lr){8-10}
\textbf{Task} & \textbf{\#Files} & \textbf{\#LoC} &
\textbf{\#Num} &
\textbf{\#Mod} & \textbf{\#Edges} & \textbf{Depth} &
\textbf{Avg.\ LoC} & \textbf{Max\ LoC} & \textbf{Width} \\
\midrule
task\_ClickHouse\_seg07 & 3.3k & 511k & 817 & 30 & 2121 & 34 & 625 & 71.4k & 404 \\
task\_NodeBB\_seg01 & 1.7k & 396k & 138 & 17 & 640 & 27 & 2.9k & 64.2k & 28 \\
task\_NodeBB\_seg03 & 4.1k & 294k & 382 & 22 & 130 & 10 & 770 & 35.5k & 166 \\
task\_NodeBB\_seg05 & 4.6k & 130k & 355 & 20 & 127 & 10 & 365 & 11.3k & 310 \\
task\_TypeScript\_seg01 & 14.4k & 379k & 406 & 18 & 1119 & 47 & 934 & 95.6k & 304 \\
task\_afl\_fuzzing & 11 & 5142 & 5 & 5 & 4 & 2 & 1.0k & 1400 & 4 \\
task\_alloy\_formal\_modeling & 18 & 6457 & 4 & 4 & 4 & 4 & 1.6k & 1845 & 1 \\
task\_angr\_concolic\_execution & 14 & 5963 & 5 & 5 & 5 & 3 & 1.2k & 1295 & 3 \\
task\_ansible\_seg11 & 5.1k & 45.8k & 319 & 20 & 704 & 17 & 144 & 11.8k & 134 \\
task\_ansible\_seg13 & 5.8k & 25.6k & 111 & 16 & 59 & 5 & 231 & 6010 & 69 \\
task\_architecture\_riscv\_labs & 420 & 1130 & 15 & 15 & 7 & 3 & 75.3 & 273 & 9 \\
task\_boogie\_dafny & 21 & 7025 & 3 & 3 & 2 & 3 & 2.3k & 2811 & 1 \\
task\_calcite\_query\_optimization & 9 & 6411 & 3 & 3 & 2 & 2 & 2.1k & 2741 & 2 \\
task\_cocos2dx\_navmesh\_medium & 4.5k & 5207 & 14 & 14 & 12 & 4 & 372 & 1143 & 9 \\
task\_cocos2dx\_physics2d\_medium & 4.2k & 4700 & 18 & 18 & 21 & 7 & 261 & 503 & 9 \\
task\_compiler & 11 & 1057 & 27 & 27 & 31 & 8 & 39.1 & 107 & 6 \\
task\_compiler\_cmm\_mips & 29 & 2688 & 9 & 9 & 7 & 4 & 299 & 1839 & 6 \\
task\_compiler\_fdmj\_llvm & 89 & 1304 & 7 & 7 & 5 & 4 & 186 & 472 & 2 \\
task\_compiler\_rustlike\_rv32 & 97 & 2079 & 6 & 6 & 6 & 4 & 346 & 606 & 2 \\
task\_compiler\_sysy\_rust & 21 & 1906 & 6 & 6 & 5 & 4 & 318 & 591 & 2 \\
task\_cpachecker & 26 & 80 & 3 & 3 & 2 & 3 & 26.7 & 34 & 1 \\
task\_cs61b\_extra\_java\_bundle & 218 & 2837 & 22 & 22 & 13 & 3 & 129 & 308 & 10 \\
task\_cvc\_smt\_solver & 31 & 8013 & 3 & 3 & 2 & 3 & 2.7k & 3065 & 1 \\
task\_datastructs\_java & 219 & 1977 & 15 & 15 & 4 & 2 & 132 & 330 & 11 \\
task\_db\_nanodb & 417 & 1522 & 9 & 9 & 8 & 3 & 169 & 774 & 4 \\
task\_db\_query\_optimizer\_labs & 70 & 1005 & 26 & 26 & 41 & 26 & 38.7 & 111 & 1 \\
task\_db\_storage\_index\_labs & 240 & 2398 & 10 & 10 & 11 & 7 & 240 & 1374 & 3 \\
task\_dbcompiler & 21 & 1217 & 24 & 24 & 21 & 10 & 50.7 & 161 & 5 \\
task\_dcqcn\_rdma\_evolution & 2.3k & 31.2k & 4 & 11 & 3 & 4 & 7.8k & 15.7k & 1 \\
task\_deep\_learning\_three\_assignments & 91 & 1365 & 16 & 16 & 10 & 4 & 85.3 & 202 & 7 \\
task\_deep\_rl\_coursework & 203 & 3285 & 19 & 19 & 23 & 6 & 173 & 811 & 6 \\
task\_dist\_sys\_raft\_go & 89 & 1173 & 4 & 4 & 5 & 3 & 293 & 581 & 2 \\
task\_distributed\_fs\_chfs & 2.1k & 2875 & 12 & 12 & 15 & 7 & 240 & 1195 & 3 \\
task\_django\_seg01 & 6.2k & 16.5k & 209 & 16 & 476 & 21 & 78.8 & 1687 & 45 \\
task\_django\_seg05 & 6.6k & 15.9k & 221 & 16 & 357 & 15 & 71.8 & 1690 & 108 \\
task\_django\_seg11 & 6.8k & 10.6k & 165 & 17 & 213 & 12 & 64.0 & 718 & 80 \\
task\_doop\_souffle\_static\_analysis & 17 & 6342 & 5 & 5 & 4 & 3 & 1.3k & 1458 & 2 \\
task\_ds\_five\_assignments & 27 & 4914 & 5 & 5 & 0 & 1 & 983 & 1231 & 5 \\
task\_echarts\_seg01 & 1.3k & 19.9k & 52 & 18 & 140 & 14 & 382 & 3903 & 15 \\
task\_echarts\_seg04 & 1.4k & 26.3k & 53 & 20 & 153 & 13 & 496 & 12.6k & 22 \\
task\_echarts\_seg05 & 1.5k & 14.2k & 43 & 20 & 140 & 12 & 330 & 3271 & 20 \\
task\_echarts\_seg07 & 1.5k & 9980 & 65 & 10 & 223 & 13 & 154 & 2093 & 29 \\
task\_echarts\_seg08 & 1.6k & 28.3k & 77 & 11 & 340 & 20 & 367 & 4742 & 32 \\
task\_echarts\_seg09 & 1.6k & 6779 & 46 & 11 & 128 & 12 & 147 & 1037 & 26 \\
task\_echarts\_seg10 & 1.6k & 17.3k & 47 & 24 & 72 & 9 & 367 & 2229 & 30 \\
task\_echarts\_seg11 & 1.6k & 12.4k & 62 & 38 & 215 & 17 & 200 & 1679 & 31 \\
task\_echarts\_seg13 & 1.9k & 6123 & 46 & 34 & 60 & 7 & 133 & 1509 & 25 \\
task\_evosuite\_test\_generation & 17 & 8216 & 6 & 6 & 6 & 3 & 1.4k & 1434 & 3 \\
task\_frama\_c & 21 & 4659 & 3 & 3 & 2 & 3 & 1.6k & 2095 & 1 \\
task\_framework\_seg04 & 1.4k & 130k & 383 & 18 & 766 & 22 & 339 & 15.6k & 147 \\
task\_gem5\_architecture\_simulation & 4 & 6004 & 4 & 4 & 3 & 2 & 1.5k & 1661 & 3 \\
task\_godot\_bootstrap\_medium & 13.8k & 5020 & 14 & 14 & 15 & 8 & 359 & 1394 & 3 \\
task\_godot\_navigation2d\_medium & 13.8k & 5239 & 11 & 11 & 12 & 5 & 476 & 1261 & 7 \\
task\_graph\_processing & 6 & 5399 & 3 & 3 & 2 & 3 & 1.8k & 1819 & 1 \\
task\_graphics\_raster\_2d & 17 & 1305 & 12 & 12 & 17 & 4 & 108.8 & 529 & 7 \\
task\_graphics\_scotty3d & 390 & 1199 & 8 & 8 & 2 & 3 & 150 & 276 & 6 \\
\bottomrule
\end{tabular}
}
\end{table}

\begin{table}[!htbp]
\centering
\caption{Structural statistics for \lhb{} tasks, Part 2.}
\label{tab:lhb_task_structure_stats_part2}
\scriptsize
\setlength{\tabcolsep}{3.0pt}
\renewcommand{\arraystretch}{1.05}
\resizebox{\textwidth}{!}{%
\begin{tabular}{@{}lrr|r|rrr|rrr@{}}
\toprule
\multicolumn{3}{c|}{\textbf{Repository}} &
\multicolumn{1}{c|}{\textbf{Units}} &
\multicolumn{3}{c|}{\textbf{Task structure}} &
\multicolumn{3}{c}{\textbf{Implementation target}} \\
\cmidrule(lr){1-3}
\cmidrule(lr){4-4}
\cmidrule(lr){5-7}
\cmidrule(lr){8-10}
\textbf{Task} & \textbf{\#Files} & \textbf{\#LoC} &
\textbf{\#Num} &
\textbf{\#Mod} & \textbf{\#Edges} & \textbf{Depth} &
\textbf{Avg.\ LoC} & \textbf{Max\ LoC} & \textbf{Width} \\
\midrule
task\_hadoop\_seg05 & 36.3k & 533k & 318 & 16 & 597 & 16 & 1.7k & 103k & 142 \\
task\_java\_ds\_coursework & 96 & 3997 & 17 & 17 & 10 & 4 & 235 & 522 & 6 \\
task\_jenkins\_seg02 & 23.6k & 494k & 84 & 11 & 167 & 16 & 5.9k & 43.5k & 40 \\
task\_jenkins\_seg05 & 12.3k & 63.4k & 44 & 7 & 40 & 7 & 1.4k & 13.1k & 21 \\
task\_jpf\_symbolic\_execution & 1.2k & 229k & 3 & 3 & 2 & 3 & 76.4k & 126k & 1 \\
task\_ligra\_graph\_evolution & 57 & 14.8k & 2 & 8 & 1 & 2 & 7.4k & 7990 & 1 \\
task\_linux011\_kernel\_hacks & 137 & 2690 & 13 & 13 & 15 & 6 & 207 & 724 & 5 \\
task\_llvm\_alive\_verification & 556 & 32.8k & 3 & 3 & 2 & 2 & 10.9k & 27.0k & 2 \\
task\_lsm\_vector\_store\_four\_stage & 1.2k & 1841 & 6 & 6 & 6 & 4 & 307 & 1832 & 2 \\
task\_mininet\_network\_simulation & 10 & 6240 & 5 & 5 & 5 & 3 & 1.2k & 1339 & 3 \\
task\_ml\_four\_assignments & 42 & 2002 & 10 & 10 & 7 & 3 & 200 & 651 & 5 \\
task\_mlir\_circuit\_ir & 10 & 5037 & 3 & 3 & 2 & 3 & 1.7k & 2237 & 1 \\
task\_mlp\_numpy\_coursework & 64 & 1024 & 7 & 7 & 6 & 3 & 146 & 329 & 4 \\
task\_mlsys\_autodiff\_pa1 & 18 & 1263 & 20 & 20 & 34 & 10 & 63.1 & 360 & 3 \\
task\_monogame\_math2d\_medium & 2.1k & 5002 & 12 & 12 & 17 & 4 & 417 & 1393 & 4 \\
task\_monogame\_transform3d\_medium & 2.1k & 4481 & 8 & 8 & 12 & 4 & 560 & 2318 & 2 \\
task\_mysql\_sql\_labs & 48 & 1093 & 9 & 9 & 7 & 3 & 121 & 413 & 5 \\
task\_nandteris & 256 & 1020 & 6 & 6 & 8 & 5 & 170 & 278 & 2 \\
task\_navidrome\_seg02 & 364 & 52.1k & 64 & 11 & 9 & 3 & 814 & 11.0k & 49 \\
task\_navidrome\_seg04 & 570 & 29.1k & 56 & 12 & 16 & 5 & 520 & 13.1k & 44 \\
task\_navidrome\_seg05 & 627 & 50.1k & 80 & 12 & 14 & 4 & 626 & 17.4k & 59 \\
task\_navidrome\_seg07 & 752 & 7595 & 44 & 12 & 4 & 3 & 173 & 1448 & 37 \\
task\_net\_os\_labs & 54 & 2019 & 18 & 18 & 10 & 6 & 112 & 146 & 9 \\
task\_networks\_proxy\_rdt & 48 & 1007 & 22 & 22 & 15 & 3 & 45.8 & 216 & 14 \\
task\_node\_seg03 & 38.8k & 126k & 321 & 24 & 358 & 15 & 393 & 49.6k & 210 \\
task\_ns3\_network\_simulation & 3.2k & 337k & 2 & 2 & 1 & 2 & 168.3k & 190k & 1 \\
task\_numpy\_dl\_aics\_style & 176 & 1001 & 20 & 20 & 23 & 12 & 50.0 & 113 & 3 \\
task\_opengl\_glfw\_labs & 1.4k & 1259 & 8 & 8 & 7 & 8 & 157 & 274 & 1 \\
task\_os\_c\_fs\_labs & 28 & 1230 & 7 & 7 & 1 & 2 & 176 & 639 & 6 \\
task\_os\_distributed\_practicals & 19 & 1049 & 5 & 5 & 0 & 1 & 210 & 595 & 5 \\
task\_os\_labs\_asm\_kernel & 780 & 14.3k & 11 & 11 & 9 & 5 & 1.3k & 3874 & 4 \\
task\_panda3d\_collision\_hard & 6.7k & 4813 & 14 & 14 & 14 & 5 & 344 & 903 & 6 \\
task\_panda3d\_pgraph\_medium & 6.6k & 3767 & 12 & 12 & 11 & 4 & 314 & 895 & 5 \\
task\_perses\_program\_reduction & 21 & 5853 & 5 & 5 & 5 & 3 & 1.2k & 1283 & 3 \\
task\_rails\_seg09 & 4.8k & 11.4k & 122 & 16 & 90 & 7 & 93.7 & 1146 & 78 \\
task\_raylib\_shapes2d\_medium & 1.4k & 5083 & 10 & 10 & 18 & 5 & 508 & 1491 & 5 \\
task\_raylib\_textures\_hard & 1.4k & 6359 & 11 & 11 & 26 & 5 & 578 & 1548 & 5 \\
task\_rdma\_ns3\_evolution & 2.3k & 590 & 3 & 12 & 2 & 3 & 197 & 527 & 1 \\
task\_riscv\_cpu\_six\_labs & 305 & 3429 & 6 & 6 & 6 & 3 & 572 & 720 & 4 \\
task\_riscv\_cpu\_sv & 133 & 1875 & 12 & 12 & 17 & 6 & 156 & 449 & 5 \\
task\_riscv\_cpu\_verilog\_three & 56 & 1157 & 6 & 6 & 6 & 5 & 193 & 275 & 2 \\
task\_riscv\_os\_multiproject & 619 & 4254 & 7 & 7 & 5 & 6 & 608 & 2006 & 2 \\
task\_sat\_minisat\_glucose & 29 & 17.3k & 2 & 9 & 1 & 2 & 8.7k & 16.9k & 1 \\
task\_security\_labs & 84 & 1056 & 25 & 25 & 23 & 6 & 42.2 & 103 & 7 \\
task\_sicp\_py\_scheme\_sql & 696 & 3135 & 22 & 22 & 25 & 13 & 142 & 698 & 5 \\
task\_sklearn\_ml\_toolchain & 4 & 5252 & 3 & 3 & 3 & 3 & 1.8k & 2445 & 1 \\
task\_soot\_taint\_analysis & 10 & 5212 & 3 & 3 & 2 & 3 & 1.7k & 1928 & 1 \\
task\_spring\_course\_mgmt\_fullstack & 19.8k & 1116 & 10 & 10 & 15 & 4 & 112 & 263 & 3 \\
task\_sql\_engine\_myjql & 44 & 1263 & 5 & 5 & 5 & 4 & 253 & 524 & 2 \\
task\_stream\_mining & 5 & 5245 & 5 & 5 & 5 & 3 & 1.0k & 1241 & 2 \\
task\_systems\_csapp\_labs & 101 & 1289 & 6 & 6 & 1 & 2 & 215 & 323 & 5 \\
task\_systems\_four\_labs & 342 & 1913 & 10 & 10 & 5 & 3 & 191 & 623 & 6 \\
task\_tcp\_course\_stack & 48 & 1405 & 2 & 2 & 1 & 2 & 702 & 1192 & 1 \\
task\_x86\_os\_four\_labs & 128 & 2229 & 9 & 9 & 11 & 6 & 248 & 600 & 2 \\
task\_xjqkl & 26 & 1307 & 6 & 6 & 9 & 4 & 218 & 632 & 3 \\
task\_z3\_fixedpoint\_verification & 11 & 6074 & 5 & 5 & 5 & 4 & 1.2k & 1345 & 2 \\
\bottomrule
\end{tabular}
}
\end{table}

\FloatBarrier
\clearpage
\section{Task Schema for Loop Engineering Evaluation}\label{app:task-schema}

\includegraphics[width=\linewidth]{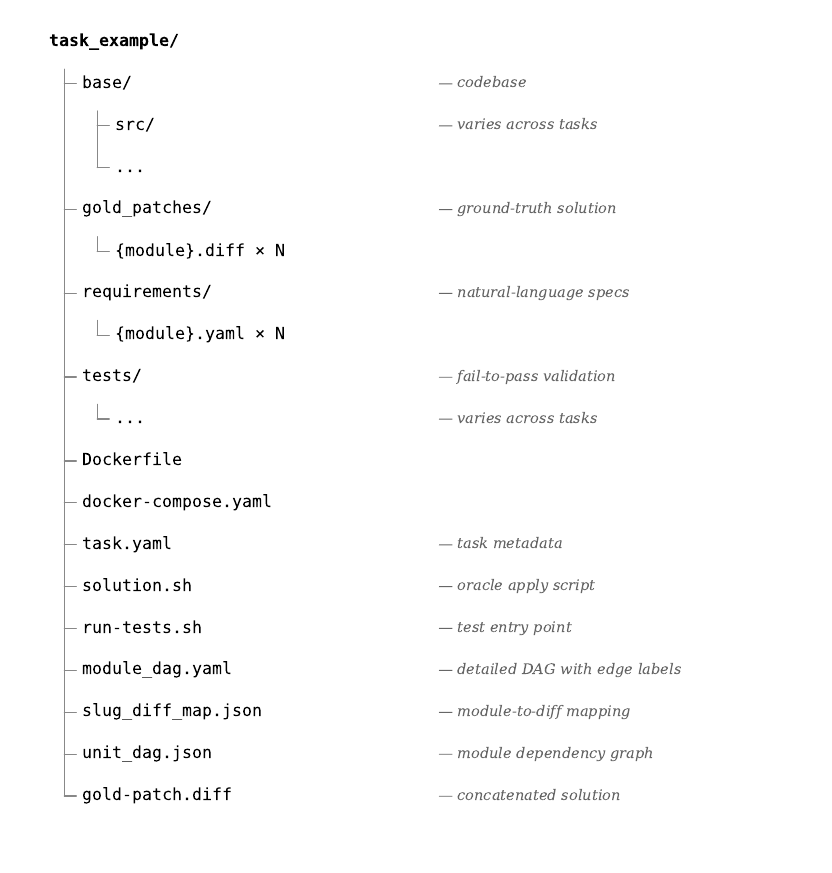}

\begin{figure}[!t]
\centering
\definecolor{schemaframe}{HTML}{4A6FA5}
\definecolor{schemabody}{HTML}{F2F5FB}
\definecolor{schemanote}{HTML}{555555}
\begin{tcolorbox}[
  enhanced, sharp corners=south,
  colback=schemabody, colframe=schemaframe,
  coltitle=white, fonttitle=\small\bfseries\sffamily,
  title={On-disk layout of a single \lhb{} task},
  left=6pt, right=6pt, top=4pt, bottom=4pt,
  boxrule=0.5pt, titlerule=0pt,
]
\footnotesize
\newcommand{\tn}[1]{{\rmfamily\color{schemanote}\textit{#1}}}
\begin{tabularx}{\linewidth}{@{}>{\ttfamily}l@{\hspace{8pt}}>{\raggedright\arraybackslash}X@{}}
\textbf{\textless repo\textgreater{}-\textless task-id\textgreater/} & \tn{self-contained task bundle} \\
\quad Dockerfile & \tn{pinned base image, system deps, lockfile} \\
\quad docker-compose.yaml & \tn{tasks with multiple services} \\
\quad run-tests.sh & \tn{checkpoint test entry point for the evaluation runtime} \\
\quad solution.sh & \tn{driver that applies the reference solution} \\
\quad gold-patch.diff & \tn{reference human solution as a unified diff} \\
\quad task.yaml & \tn{global task metadata and DAG topology} \\
\quad units/ & \tn{requirements and test bindings presented during evaluation} \\
\quad\quad unit\_001.yaml & \tn{requirement, scope, parents, attached tests} \\
\quad\quad unit\_002.yaml & \\
\quad\quad \dots & \\
\quad\quad unit\_NNN.yaml & \\
\quad tests/ & \tn{source-derived and supplemented test files} \\
\quad\quad unit\_001/ & \tn{tests adjudicating unit 001} \\
\quad\quad unit\_002/ & \\
\quad\quad \dots & \\
\quad\quad unit\_NNN/ & \\
\quad base/ & \tn{entry-commit codebase the patch is applied to} \\
\end{tabularx}
\end{tcolorbox}
\vspace{-6pt}
\caption{Schema of a materialized \lhb{} task.}
\label{fig:task-schema}
\end{figure}

Each task is materialized as a self contained directory for loop engineering evaluation, separating the work graph, presented requirements, executable obligations, and reference solution. This separation maintains a distinction between the evaluated loop's input contract and the hidden artifacts used to validate obligations and compute loop trace metrics.

The \texttt{Dockerfile} and optional \texttt{docker-compose.yaml} pin a reproducible execution environment.

\texttt{run-tests.sh} is the checkpoint test entry point used by the evaluation runtime to adjudicate the current workspace state.

\texttt{solution.sh} and \texttt{gold-patch.diff} store the reference human solution used to validate executable obligations and anchor implementation comparisons.

\texttt{task.yaml} declares task metadata and per unit fields, including requirement text, file and symbol scope, dependency parents, and attached tests.

The \texttt{tests/} subtree holds source derived and supplemented acceptance tests indexed to the unit they adjudicate.

The evaluation contract is uniform across the three task sources and separates the full requirement set presented to the evaluated loop from the active obligations counted at each checkpoint. The evaluation runtime mounts the task directory into the container, presents \emph{all} per unit requirement manifests up front, invokes \texttt{run-tests.sh} at committed checkpoints, and records the resulting test state vector $s_i^{(\tau)}$ used to compute the loop trace metrics in Appendix~\ref{app:metrics}. The \emph{ready frontier} controls newly released obligations, while previously passed released tests remain active as regression obligations during later work.

\begin{table}[t]
  \centering
  \caption{\rev{Visibility contract used by \lhb{}. Unit manifests and attached tests are presented up front, while checkpoint progress and scoring state remain evaluator only.}}
  \label{tab:visibility-contract}
  \small
  \setlength{\tabcolsep}{5pt}
  \renewcommand{\arraystretch}{1.12}
  \begin{tabularx}{\linewidth}{@{}>{\raggedright\arraybackslash}X >{\raggedright\arraybackslash}X@{}}
    \toprule
    \textbf{Evaluated-loop visible} & \textbf{Evaluator only} \\
    \midrule
    Global task instruction and unit manifests, including requirements, scopes, declared parents, and attached test bindings & Active ready frontier and newly released obligation state \\
    Base repository, attached and project native tests, and outputs of commands initiated by the evaluated loop & Reference patches, gold solution, reference implementation, and offline construction transcripts \\
    Tests authored and run by the evaluated loop & Checkpoint outcomes, accumulated regression obligations, and scoring state \\
    \bottomrule
  \end{tabularx}
\end{table}
\section{Materialized Task Example for Loop Engineering Evaluation}\label{app:example-task}

This appendix presents one materialized task and the loop trace metrics that \lhb{} records during evaluation, making the schema in Section~\ref{sec:design} and Appendix~\ref{app:task-schema} concrete. The example is a PR Sequences task drawn from an actively maintained open source repository. It connects materialization steps to loop trace metrics, including commit history snapshotting, per pull request diff extraction by role, dependency routing, ready frontier test release, and regression obligation retention. Under the contamination protocol of Appendix~\ref{app:contamination}, identifiable tokens are redacted and marked as \texttt{<repo>}, \texttt{<unit-id>}, and \texttt{<author>}.

\paragraph{Native artifact.}

The task originates from a sequence of \textbf{9} merged pull requests against \texttt{<repo>} over \textbf{4.2 months}, touching \textbf{37} source files and \textbf{14} test files. The base codebase at the entry commit comprises \textbf{18.4k} LoC across \textbf{142} files. Each pull request provides a description, a per file diff, and a CI run record from the upstream test suite, giving source evidence for implementation units, prerequisite relations, and executable obligations.

\paragraph{Preprocessing output.}

After preprocessing, the nine pull requests decompose into \textbf{12} atomic candidate units. Three original PRs are split when their diffs touch disjoint feature surfaces, and two adjacent PRs are merged when the second repairs a regression introduced by the first and is therefore dependent on that PR. The materialization extractor produces the aggregated segments under the prescribed extraction procedure, and the segments pass the fixed acceptance standard on the first attempt.

\paragraph{Materialized task bundle.}
The candidate is rewritten into the uniform schema of Appendix~\ref{app:task-schema}. The resulting on disk layout is:
\begin{promptbox}{Task bundle layout}
<repo>-<unit-id>/
  Dockerfile
  docker-compose.yaml
  run-tests.sh
  solution.sh
  gold-patch.diff
  task.yaml
  tests/
    unit_001/...
    unit_002/...
    ...
    unit_012/...
\end{promptbox}

The \texttt{Dockerfile} pins Python 3.11.6, three system packages absent from the upstream README but recovered from CI configuration, and a lockfile derived from the entry commit. The materialization stage verifies \texttt{gold-patch.diff} successfully after a single build attempt.

\paragraph{Per unit requirement.}
A representative unit's \texttt{task.yaml} entry (paraphrased and redacted) reads:
\begin{promptbox}{Per unit requirement (excerpt)}
unit_id: <unit-id>
requirement: |
  Extend the streaming response handler so that
  partial chunks emitted before a terminating
  sentinel are buffered into a single payload
  delivered to downstream consumers, preserving
  ordering across concurrent producers.
file_scope:
  - src/stream/handler.py
  - src/stream/buffer.py
symbol_scope:
  - StreamHandler.consume
  - StreamHandler._flush
acceptance:
  - concurrent producers preserve emission order
  - sentinel terminates buffering atomically
  - downstream observers see one payload per
    logical message
parents: [unit_003, unit_005]
tests:
  - tests/unit_007/test_ordering.py
  - tests/unit_007/test_sentinel.py
  - tests/unit_007/test_concurrent.py
\end{promptbox}

The requirement is phrased as an acceptance contract. Procedural directives, command level instructions, and references to the gold patch are omitted. The resulting loop trace records whether the evaluated loop infers ordering and integration constraints from the dependency graph and released tests without access to the hidden implementation path.

\paragraph{Recovered DAG.}

The recovered DAG encodes the routing and obligation structure recorded by the evaluation runtime. DAG recovery (Section~\ref{sec:dag-framework}) admits \textbf{18} edges over the 12 units under the unambiguous evidence rules. The longest dependency chain has depth \textbf{6}. The maximum out degree is \textbf{4}, corresponding to a foundational refactor unit imported by later units. Three units form an independent branch with zero incoming edges from the rest of the task, creating a parallel ready frontier midway through the PR sequence.

\paragraph{Trial verification.}
The materialized test suite passes the solvability, non-triviality, and discriminativeness trials of Section~\ref{sec:description-refine} on the first run for \textbf{10} of \textbf{12} units. Two units undergo one repair round during materialization. In one case a flaky timing assertion is replaced with a deterministic state check. In the other, a discriminativeness trial reveals that two predecessor tests already cover the unit's full contract, and the materialization pipeline strengthens the unit suite with an additional assertion targeting the new contract.

\paragraph{Evaluation trace excerpt.}

Under the protocol of Section~\ref{sec:protocol}, the recorded loop trace contains \textbf{31} checkpoints over the evaluation episode. The visitation sequence $\sigma$ differs from the human order $\sigma^*$ on \textbf{3} of 12 units, all within branches with partially ordered alternatives under the DAG. Two regressions are recorded against previously completed units, and one is recovered before the episode ends. The full test state matrix $S$ is omitted for brevity, while Appendix~\ref{app:metrics} defines the corresponding loop trace metrics.
\FloatBarrier
\section{Contamination Checks for Loop Engineering Evaluation}\label{app:contamination}

\lhb{} derives its loop engineering evaluation tasks from public development artifacts, including PR Sequences, Course Labs, and Research Evolutions. This source choice creates potential overlap with modern LLM pretraining data. We therefore separate source recognition risk from the loop trace metrics used to evaluate loop execution.

\paragraph{Input Side Anonymization.}
During Task Instrumentation (Section~\ref{sec:description-refine}), every unit requirement is rewritten to remove identifiable tokens that could let a model recognize the source from the prompt: repository names, course identifiers (e.g., MIT 6.824, Stanford CS144), paper titles, and author attributions. For PR Sequences, the same rewriting policy is applied at the segment (SEG) level, where original commit messages and PR titles often include the feature name, issue tracker ID, or merging author. After this step, the task prompt provided to the evaluated loop contains the acceptance specification, public interfaces, and tests. Source identifying evidence that would support retrieval by name is removed before evaluation.

\paragraph{Memorization and Loop Trace Metrics.}

Anonymization addresses prompt level identification, but training set memorization may remain in model weights. A model that has internalized a reference implementation, for example, may reproduce local algorithmic fragments regardless of prompt wording. Two observations limit the extent to which this residual risk explains the loop trace metrics recorded by \lhb{}.

The first observation is empirical. The strongest configuration in our evaluation, Claude Code with Opus-4.7, uses the native Anthropic model and loop infrastructure under the most permissive context budget and reaches a Resolve Rate of $25.00\%$ (Table~\ref{tab:rq1}). If weight level memorization were the principal driver, frontier model and loop configurations would be expected to approach higher resolution on canonical Course Labs and Research Evolutions tasks, whose source material is publicly accessible. This result provides limited support for memorization as the main explanation.

The second observation is structural. \lhb{} treats recall of a single implementation fragment as insufficient evidence for loop engineering evaluation. A resolved task entails prerequisite consistency across many edits, retained obligations as the codebase evolves, preserved state across context renewal, and test feedback routed along the dependency DAG. These loop conditions extend beyond pretraining exposure to a solution for a single local unit. Memorization may supply fragments, but \lhb{} records whether those fragments remain consistent under live regression obligations, where current model and loop configurations exhibit constrained state continuity and obligation retention.

\paragraph{Model cutoff robustness slice.}

A direct slice by source ingest date and model cutoff can partition tasks by whether their source material predates each model's training cutoff and report Resolve Rate within each partition. The current analysis reports the aggregate constraint that the strongest model and loop configuration reaches $25\%$ Resolve Rate, providing limited support for memorization driven inflation as the principal explanation.
\section{Evaluation Settings for Model and Loop Configurations}\label{app:exp-settings}

\paragraph{Models and provider routes.}
RQ1 (E1) isolates model choice by comparing eight model checkpoints under a fixed Claude Code loop implementation: Opus-4.7, GPT-5.5, GLM-5.1, DeepSeek-V4P, Gemini-3.1-Pro, Qwen3.6-Plus, Kimi-2.6, and Grok-4.1-FR.
RQ1 (E2) isolates loop implementation by fixing the model to \texttt{azure/gpt-5.4-20260305} and varying Claude Code, Codex, GitHub Copilot, OpenHands, SWE-agent, and mini-SWE-agent.
RQ1 (E3) examines within family model changes while pairing each family with its corresponding loop implementation for GPT, Claude Opus, and Qwen.
Provider access is normalized through routes implemented in \texttt{long\_horizon\_bench/agents/}, including an Azure OpenAI deployment with AD token auth (\texttt{api-version} \texttt{2025-04-01-preview}), the GitHub Copilot token exchange endpoint, a DashScope Anthropic compatible proxy, and direct Anthropic SDK access for Anthropic native models. This normalization separates provider transport from the loop engineering factors attributed to model choice and loop implementation.

\paragraph{Decoding and evaluation input.}
Each loop implementation inherits provider defaults unless its evaluation adapter records an explicit override. For the Anthropic SDK path used by Claude Code, we set \texttt{max\_tokens=128{,}000} and leave temperature and top p at their SDK defaults. The SWE-agent LiteLLM call uses \texttt{temperature=0}. Top p is disabled because Anthropic rejects requests that set both fields. mini-SWE-agent forwards \texttt{model.model\_kwargs.reasoning\_effort=high} to providers that accept it. Loop implementations receive the evaluation contract in Appendix~\ref{app:prompts}; no additional custom system prompt is supplied unless the evaluation adapter defines one.

\paragraph{Runtime budget and concurrency.}
The evaluation runtime imposes a per task wall clock budget $T_{\max}$ chosen per task ($\geq$2~h, capped at 24~h for the longest PR Sequence and Research Evolution tasks) and passes it to each loop implementation as \texttt{timeout\_sec}.
For loop implementations with their own iteration cap, we set the cap above the wall clock budget, making wall clock time the evaluation horizon. Claude Code SDK uses \texttt{max\_iterations=10{,}000}, and mini-SWE-agent uses \texttt{--cost-limit 0} (uncapped) with \texttt{--exit-immediately}.
Each bash tool call inside the container is independently capped at 600~s, separating local command stalls from sustained loop progress.
Trials run concurrently in a \texttt{ThreadPoolExecutor}. Per trial credentials are passed as explicit kwargs through the task runner, keeping cross trial authorization races outside attribution to the evaluated loop implementation.
\section{Metrics for Loop Engineering Evaluation}\label{app:metrics}

We formalize the five headline RQ1 metrics used to evaluate model and loop configurations on \lhb{}. Together, they separate end state scoring from loop trace metrics for dependency progress, obligation retention, and resource use. Let a task have tests indexed by $i$ and evaluation checkpoints indexed by $\tau \in \{1,\dots,T\}$. Let $s_i^{(\tau)} \in \{0,1\}$ denote the pass/fail outcome of test $i$ at checkpoint $\tau$, and let $O_\tau$ denote the set of \emph{regression obligations} active at checkpoint $\tau$ (tests previously released and passed at some earlier checkpoint).
The recorded test state matrix is $S = [s_i^{(\tau)}]$ and serves as the evidence basis from which both final outcomes and intermediate loop trace metrics are computed.
Let $D_\tau$ denote the set of tests released by checkpoint $\tau$, and let $T_{\max}$ denote the wall clock time budget. The released test pass rate at checkpoint $\tau$ is
\begin{equation}
\operatorname{TPR}(\tau) \;=\; \frac{\bigl|\{\, i \in D_\tau : s_i^{(\tau)} = 1 \,\}\bigr|}{|D_\tau|}.
\end{equation}

\subsection{Resolve Rate (RR)}\label{app:final-success}
\begin{equation}
\operatorname{RR} \;=\; \mathbb{1}\!\left[\,\bigwedge_{i \in \text{all tests}} s_i^{(T)} = 1\,\right].
\end{equation}
RR records whether every released test passes at task end, including regression obligations accumulated as tests are released. It is the strict end state outcome metric, while the remaining metrics record dependency progress, obligation retention, and resource use along the recorded loop trace.

\subsection{Test Pass Rate (TPR)}\label{app:pass-rate}
\begin{equation}
\operatorname{TPR} \;=\; \frac{1}{|D_T|}\sum_{i \in D_T} s_i^{(T)}.
\end{equation}
TPR records the average fraction of released tests passing at task end. Unlike RR, it preserves partial progress over the released test set and connects the final state to the intermediate loop trace recorded during execution.

\subsection{Dependency Depth (Depth)}\label{app:depth}

Depth measures dependency reach before the first residual handoff by the outer evaluation loop, isolating how far one continuous execution segment preserves prerequisite obligations before residual continuation begins. Let each released test $i$ inherit the topological layer of the development unit it adjudicates, $\mathrm{layer}(i) \in \{0, \dots, L\}$, where $L = \max_{v \in V} \mathrm{layer}(v)$ is the longest prerequisite chain of the task DAG. Let $\tau^{\dagger}$ denote the last checkpoint of the first outer evaluation round, that is, the final snapshot before the outer loop, if any, restarts the loop with a freshly derived residual. Let $P_{\tau^{\dagger}} = \{i : s_i^{(\tau^{\dagger})} = 1\}$ denote the tests passing at $\tau^{\dagger}$. Depth is the deepest layer at which every prerequisite test still passes at $\tau^{\dagger}$, normalized by $L$:
\begin{equation}
\resizebox{.9\columnwidth}{!}{$\displaystyle
\operatorname{Depth} \;=\; \mathbb{E}_{\text{task}}\!\left[\,\frac{\max\bigl\{\ell : \forall i,\, \mathrm{layer}(i) \leq \ell \Rightarrow i \in P_{\tau^{\dagger}}\bigr\}}{L}\,\right].
$}
\end{equation}
Depth records dependency reach before the first residual handoff in the evaluated loop, while RR combines this reach with residual recovery across later rounds. Resolved tasks may reach $\mathrm{RR}=1$ through high Depth before the first residual handoff or through lower initial Depth followed by successful residual recovery. Read together, the two metrics separate entry layer stalls ($\operatorname{Depth}{\ll}1$) from late prerequisite discontinuities, where high Depth pairs with low RR after later execution no longer preserves prior obligations.

\subsection{Regression Rate (Reg)}\label{app:regression-rate}

A \emph{regression event} occurs whenever a previously passing obligation test fails at a later evaluation point, making obligation retention observable as a loop trace metric rather than solely as final nonresolution. Checkpoints $\tau$ are snapshots emitted by the dual container watcher (Section~\ref{sec:protocol}). When the cumulative diff from the evaluated loop exceeds a fixed line threshold against the snapshot baseline, the watcher container freezes the workspace and runs the full released test suite, recording per test pass or fail outcomes in \texttt{observation\_history.jsonl} alongside the \texttt{newly\_passed} and \texttt{newly\_failed} deltas relative to the previous snapshot. Let $\tau^-$ denote the immediately preceding triggered snapshot at which test $i$ was evaluated, let $O_\tau$ denote the obligation set at $\tau$ (the union of all tests that were \texttt{newly\_passed} at any earlier triggered snapshot), and define
\begin{equation}
R = \#\bigl\{(i,\tau): i \in O_\tau,\; s_i^{(\tau^-)} \!=\! 1,\; s_i^{(\tau)} \!=\! 0\bigr\}.
\end{equation}
Then
\begin{equation}
\operatorname{Reg} \;=\; \frac{R}{\sum_\tau |O_\tau|}.
\end{equation}
Equivalently, Reg measures how often a previously satisfied obligation stops passing within the recorded loop trace, making obligation retention a directly measured loop trace metric.
Reg is interpreted jointly with Resolve Rate. Configurations with very low Resolve Rate satisfy a limited obligation set, and $|O_\tau|$ therefore provides limited exposure for estimating retention. Low Reg in that regime provides limited evidence of stable long horizon execution. Conversely, a model and loop configuration with moderate RR but high Reg, such as DeepSeek-V4P at $\text{RR}{=}11.61\%$ and $\text{Reg}{=}16.60\%$ in our runs, corresponds to obligation retention degradation beyond limited progress alone. We report Reg as a loop trace metric and use Resolve Rate for outcome ranking.

\subsection{Tokens (Tok)}\label{app:tokens}

Tok records total tokens consumed per task, summed across input and output over the full loop trace. This metric reports the resource cost of sustaining the evaluated loop over long horizon execution and contextualizes loop trace improvements against runtime expenditure.
\section{Model Metadata for Fixed Loop Comparisons}\label{app:model-release}

For the within family scaling panel in Table~\ref{tab:rq1}, we record release date and reported parameter scale for each model evaluated under a fixed loop implementation. The three families begin at different release points and follow different release cadences. The GPT family spans \texttt{gpt-4o} (May 2024) through \texttt{gpt-5.5} within the RQ1 evaluation window, the Claude family spans Sonnet-4 through Opus-4.7, and the Qwen family spans Qwen2.5-72B (open weight, $72$\,B parameters) through Qwen3.6-Plus (closed weight, undisclosed scale). Because the columns mix open and closed weight checkpoints, the metadata contextualizes fixed loop comparisons rather than defining a model capability ranking. Resolve Rate monotonicity within each family is therefore interpreted as a generational delta under a fixed loop implementation, not as a parameter count scaling law.

\begin{table}[!htbp]
  \centering
  \caption{Model metadata for RQ1 comparisons.}
  \label{tab:model-release}
  \scriptsize
  \setlength{\tabcolsep}{2pt}
  \renewcommand{\arraystretch}{1.05}
  \begin{tabular}{@{}l l l r l@{}}
    \toprule
    \textbf{Model} & \textbf{Family} & \textbf{Release} & \textbf{Params} & \textbf{Ref.} \\
    \midrule
    GPT-4o          & GPT    & 2024-05 & undisclosed & \citep{openai_gpt4o_system_card} \\
    GPT-5           & GPT    & 2025-08 & undisclosed & \citep{openai_gpt5} \\
    GPT-5.2         & GPT    & 2025-11 & undisclosed & \citep{openai_gpt52_system_card} \\
    GPT-5.4         & GPT    & 2026-03 & undisclosed & \citep{openai_gpt54} \\
    GPT-5.5         & GPT    & 2026-04 & undisclosed & \citep{openai_gpt55} \\
    \midrule
    Sonnet-4        & Claude & 2025-05 & undisclosed & \citep{anthropic_claude4_system_card} \\
    Sonnet-4.5      & Claude & 2025-09 & undisclosed & \citep{anthropic_sonnet45} \\
    Opus-4.5        & Claude & 2025-11 & undisclosed & \citep{anthropic_opus45} \\
    Opus-4.7        & Claude & 2026-02 & undisclosed & \citep{anthropic_opus47} \\
    \midrule
    Qwen2.5-72B     & Qwen   & 2024-09 & 72\,B (OW)  & \citep{qwen25_technical_report} \\
    Qwen3-Max       & Qwen   & 2025-09 & undisclosed & \citep{qwen3max} \\
    Qwen3.5-Plus    & Qwen   & 2025-12 & undisclosed & \citep{qwen35_plus} \\
    Qwen3.6-Plus    & Qwen   & 2026-03 & undisclosed & \citep{qwen36_plus} \\
    \midrule
    Gemini-3.1-Pro  & Gemini & 2026-03 & undisclosed & \citep{google_gemini31_pro} \\
    GLM-5.1         & GLM    & 2026-02 & undisclosed & \citep{zhipu_glm51} \\
    DeepSeek-V4P    & DeepSeek & 2026-01 & undisclosed & \citep{deepseek_v4_pro} \\
    Kimi-2.6        & Moonshot & 2026-03 & undisclosed & \citep{moonshot_kimi_k26} \\
    Grok-4.1-FR     & xAI    & 2026-02 & undisclosed & \citep{xai_grok41_fast_reasoning} \\
    \bottomrule
  \end{tabular}
\end{table}
\FloatBarrier
\section{Pass@K Robustness Under Fixed Loop Implementations}\label{app:passk}

The monotonic Resolve Rate trend across model generations within each family, reported in the left panel of Table~\ref{tab:rq1}, is estimated from one trial per model and loop pairing. We therefore also report pass@3 on the same model and loop pairings. The within family ordering is preserved at $K{>}1$ for every family, which supports the headline trend in Table~\ref{tab:rq1} beyond single trial variance. The limited pass@3 rate is consistent with repeated trials under a fixed model and loop configuration rarely overcoming the state continuity and regression obligation retention limitations observed during long horizon execution.

\begin{table}[h]
  \centering
  \small
  \caption{Pass@K Resolve Rate for within family scaling.}
  \label{tab:passk}
  \begin{tabular}{l cc}
    \toprule
    \textbf{Model} & \textbf{pass@1} & \textbf{pass@3} \\
    \midrule
    \multicolumn{3}{l}{\emph{GPT (Codex)}} \\
    GPT-5.5      & 21.43 & 23.21 \\
    GPT-5.2      & 13.39 & 15.18 \\
    GPT-5        & 8.04  & 9.82  \\
    GPT-4o       & 3.57  & 4.46  \\
    \midrule
    \multicolumn{3}{l}{\emph{Claude (Claude Code)}} \\
    Opus-4.7     & 25.00 & 26.79 \\
    Opus-4.5     & 17.86 & 19.64 \\
    Sonnet-4.5   & 12.50 & 14.29 \\
    Sonnet-4     & 7.14  & 8.93  \\
    \midrule
    \multicolumn{3}{l}{\emph{Qwen (Qwen Code)}} \\
    Qwen3.6-Plus & 9.82  & 11.61 \\
    Qwen3.5-Plus & 8.04  & 9.82  \\
    Qwen3-Max    & 1.79  & 2.68  \\
    Qwen2.5-72B  & 0.00  & 0.89  \\
    \bottomrule
  \end{tabular}
\end{table}
\section{Per Task Billed Tokens for RQ1}\label{app:rq1-tokens}

Tok reports mean billed tokens per task in millions under outer continuation, matched to the RQ1 configurations in Table~\ref{tab:rq1}.

\begin{table}[t]
  \centering
  \caption{Mean billed tokens per task (millions) for the RQ1 configurations.}
  \label{tab:rq1-tokens}
  \footnotesize
  \setlength{\tabcolsep}{6pt}
  \renewcommand{\arraystretch}{1.14}
  \begin{tabular}{@{}l r@{}}
    \toprule
    \textbf{Configuration} & \textbf{Tok}$\downarrow$ \\
    \midrule
    \rowcolor{macaronMintHead}\multicolumn{2}{l}{\emph{GPT (Codex)}} \\
    GPT-5.5       & 8.36M \\
    GPT-5.2       & 9.10M \\
    GPT-5         & 8.92M \\
    GPT-4o        & 4.51M \\
    \rowcolor{macaronMintHead}\multicolumn{2}{l}{\emph{Claude (Claude Code)}} \\
    Opus-4.7      & 6.91M \\
    Opus-4.5      & 5.38M \\
    Sonnet-4.5    & 4.21M \\
    Sonnet-4      & 3.76M \\
    \rowcolor{macaronMintHead}\multicolumn{2}{l}{\emph{Qwen (Qwen Code)}} \\
    Qwen3.6-Plus  & 6.96M \\
    Qwen3.5-Plus  & 6.88M \\
    Qwen3-Max     & 5.71M \\
    Qwen2.5-72B   & 5.02M \\
    \midrule
    \rowcolor{macaronPinkHead}\multicolumn{2}{l}{\textbf{Model} \textit{(fixed loop: Claude Code)}} \\
    Opus-4.7      & 6.91M \\
    GPT-5.5       & 7.18M \\
    GLM-5.1       & 4.37M \\
    DeepSeek-V4P  & 3.59M \\
    Gemini-3.1-Pro & 4.99M \\
    Qwen3.6-Plus  & 4.25M \\
    Kimi-2.6      & 2.02M \\
    Grok-4.1-FR   & 1.83M \\
    \rowcolor{macaronBlueHead}\multicolumn{2}{l}{\textbf{Loop} \textit{(fixed model: \texttt{gpt-5.4})}} \\
    Codex          & 7.94M \\
    Claude Code    & 7.03M \\
    GitHub Copilot & 8.26M \\
    OpenHands      & 6.75M \\
    SWE-agent      & 7.18M \\
    mini-swe-agent & 8.86M \\
    \bottomrule
  \end{tabular}
\end{table}
\section{Planning Loop Trace Metrics for Loop Engineering Evaluation}\label{app:planning-metrics}

We formalize the six planning metrics used in RQ2 for loop engineering evaluation. They compare the plan DAG recorded in the evaluated loop trace with the source recovered human DAG and summarize how planning choices change routing burden and obligation retention pressure (Table~\ref{tab:rq2-grid}). Let $G_A = (V_A, E_A)$ denote the recorded plan DAG and $G_H = (V_H, E_H)$ the human development DAG recovered from source records. Let $V_\cap = V_A \cap V_H$ denote the set of units present in both DAGs.

\subsection{Edge F1}\label{app:edge-f1}
Let $E_\cap = E_A \cap E_H$, and define
\begin{equation}
P_{\text{edge}} = \frac{|E_\cap|}{|E_A|}, \qquad R_{\text{edge}} = \frac{|E_\cap|}{|E_H|}.
\end{equation}
\begin{equation}
\text{Edge\,F1} \;=\; \frac{2\,P_{\text{edge}}\,R_{\text{edge}}}{P_{\text{edge}} + R_{\text{edge}}}.
\end{equation}
Edge F1 measures exact recovery of direct prerequisite edges. Low precision corresponds to unsupported dependencies in the recorded plan, while low recall corresponds to omitted prerequisites that can cause work to be routed before prerequisite state is established.

\subsection{TC-F1}\label{app:tc-f1}
Let $\operatorname{TC}(\cdot)$ denote the transitive closure of a DAG's edge set, and write $T_\cap = \operatorname{TC}(E_A) \cap \operatorname{TC}(E_H)$. Define
\begin{equation}
P_{\text{tc}} = \frac{|T_\cap|}{|\operatorname{TC}(E_A)|}, \qquad R_{\text{tc}} = \frac{|T_\cap|}{|\operatorname{TC}(E_H)|}.
\end{equation}
\begin{equation}
\text{TC\,F1} \;=\; \frac{2\,P_{\text{tc}}\,R_{\text{tc}}}{P_{\text{tc}} + R_{\text{tc}}}.
\end{equation}
TC-F1 is a more permissive measure of prerequisite ordering agreement. It credits a recorded plan that preserves \emph{$u$ before $v$} even when it uses a different intermediate path from the human development DAG, capturing whether broad routing order survives edge level mismatch.

\subsection{Layer $\rho$}\label{app:layer-rho}

Let $\ell_A(v)$ and $\ell_H(v)$ denote the topological layer index of unit $v$ in $G_A$ and $G_H$. Layer~0 contains prerequisite free nodes, and later layers depend on earlier layers.
\begin{equation}
\begin{aligned}
\operatorname{Layer}\rho \;=\; \operatorname{Spearman}\!\bigl(& \{\ell_A(v)\}_{v \in V_\cap}, \\
& \{\ell_H(v)\}_{v \in V_\cap}\bigr).
\end{aligned}
\end{equation}
Layer~$\rho$ records whether the plan places units at similar development \emph{stages} as the human reference, independent of exact edge structure. Low values correspond to the evaluated loop assigning units to different prerequisite layers, which changes when obligations become ready during execution.

\subsection{Critical Path Ratio}\label{app:cp-ratio}
Let $\operatorname{CP}(G)$ denote the length (in nodes) of the longest directed path in $G$.
\begin{equation}
\operatorname{CPR}(G) \;=\; \frac{\operatorname{CP}(G)}{|V|}.
\end{equation}
Values near~$1$ correspond to a nearly linear recorded plan that overserializes the task. Lower values correspond to preservation of parallel branches that can be routed independently. We also report a human normalized variant $\operatorname{CPR}(G_A)/\operatorname{CPR}(G_H)$ in extended results to separate task shape from loop planning behavior.

\subsection{Width Ratio}\label{app:width-ratio}
Let $w(G) = \max_\ell |\{v \in V : \ell(v) = \ell\}|$ denote the maximum topological layer width of $G$.
\begin{equation}
\operatorname{WR} \;=\; \frac{w(G_A)}{w(G_H)}.
\end{equation}
Values close to~$1$ correspond to a recorded plan that matches the human development DAG's concurrent ready frontier width. Values below~$1$ correspond to overserialization, while values above~$1$ are consistent with overparallelization caused by missing prerequisite edges and therefore higher regression obligation pressure.

\subsection{Branch Switching Rate}\label{app:branch-switch}
Assign each unit $v$ a branch identifier $b(v)$ corresponding to its earliest reachable source node in $G_H$ (units with multiple source ancestors are labeled \emph{mixed}). Given the planned or executed implementation order in the recorded loop trace $\sigma = (v_{\pi(1)}, \ldots, v_{\pi(m)})$, let $K = |\{\,i : b(v_{\pi(i)}) \neq b(v_{\pi(i+1)})\,\}|$.
\begin{equation}
\operatorname{BSR} \;=\; \frac{K}{m-1}.
\end{equation}
High values correspond to breadth first frontier coverage across branches, while low values correspond to depth first completion of one branch before switching. BSR is graph conditional and is read relative to the shape of $G_H$, since the same switching pattern can reflect frontier coverage in a wide DAG or unnecessary alternation in a narrow chain.
\section{Extended RQ2 Loop Trace Metrics}\label{app:rq2-extended}

\paragraph{Loop trace metric definitions for Table~\ref{tab:rq2-grid}.}
\emph{Planning fidelity.}
\textbf{Edge~F1} = F1 between the plan DAG edge set in the recorded loop trace and the human development DAG edge set.
\textbf{Layer~$\rho$} = Spearman correlation between topological layer indices in the recorded loop trace and those in the human development DAG. \textbf{CPR} = critical path length of $G_A$ divided by $|V_A|$ (direct value, unnormalized to gold). \textbf{WR} = mean ready frontier width of $G_A$ divided by that of $G_H$.
\emph{Implementation} metrics are computed on units resolved by the evaluated loop. \textbf{PatchLen} = patch lines per matched unit divided by gold patch lines. \textbf{Jacc} = token level Jaccard distance between candidate patches in the recorded loop trace and gold patch.
\emph{Testing} metrics summarize verification work authored during the recorded loop trace.
\textbf{\#T} = number of tests authored in the recorded loop trace per task.
\textbf{F2P} = fraction of tests authored in the recorded loop trace that fail on the initial state and pass on the gold solution.
\paragraph{Auxiliary planning metrics.}
Table~\ref{tab:rq2-grid} presents a compact metric set for planning, implementation, and testing. The auxiliary metrics here either duplicate headline columns or depend more directly on task graph shape than on loop implementation. Their appendix placement keeps the main table focused on loop trace metrics that remain comparable across loop implementations.

\paragraph{TC-F1 (planning).}
Transitive closure F1 (Appendix~\ref{app:tc-f1}) is strongly correlated with Edge F1 (Appendix~\ref{app:edge-f1}) in our runs ($\rho{>}0.9$ across the six evaluated loops), so it provides secondary evidence beyond the Edge F1 column in the main table. Its absolute level estimates whether the recorded plan captures \emph{indirect} prerequisite structure relevant to routing and obligation retention.

\paragraph{Branch Switching Rate (planning).}
BSR (Appendix~\ref{app:branch-switch}) is graph conditional. Higher BSR is consistent with frontier exploration when $G_H$ has many simultaneously ready branches, but on long sequential chains it can instead record unnecessary alternation across branches. We therefore report it as an extended loop trace metric outside the main table ranking columns.

\paragraph{Test-Ratio (testing).}
The main table reports \#T as an absolute test count. Test-Ratio equals $\#T/\#T_{\text{native}}$ with $\#T_{\text{native}}=44$, making it directly derivable from the \#T column and secondary to the reported verification effort metric.

\paragraph{JSD (testing).}
JSD measures distributional divergence between tests authored during the recorded loop trace and native test placement. Appendix placement reflects its auxiliary role relative to the main \#T and F2P columns, which more directly measure verification effort and fail to pass yield during loop engineering evaluation. Unit position histograms separate early unit concentration from late unit concentration, capturing where verification effort is routed along the horizon.

\paragraph{Source of additional patch lines.}
The main text reports that PatchLen exceeds the gold reference while Jacc to gold stays at a comparable moderate level. The recorded diffs explain the two metrics jointly. Candidate patches often impose a uniform style, add extensive comments, and rewrite existing repository blocks when repository state in the loop is incomplete. This pattern increases patch length while token level overlap remains close to the human reference, suggesting that loop state gaps can appear as implementation surplus rather than disjoint implementations.
\section{Per Source Resolve Rate for Model and Loop Configurations}\label{app:per-source}

The main experiments report aggregate Resolve Rate and loop trace metrics over the 112 \lhb{} tasks. This appendix reports the base RQ1 setting before outer continuation, matching the \textit{w/o} RR column in Table~\ref{tab:rq1}. The three task sources represent distinct evidence regimes with different routing burden, state continuity, and regression obligation retention profiles. PR Sequences inherit source evidenced dependency DAGs from merged commit history, Course Labs inherit module level dependency structure from curriculum design, and Research Evolutions inherit coarser successor chains from citation graphs. Per source reporting separates these loop engineering patterns before aggregation.
Table~\ref{tab:per-source-rr} reports per source Resolve Rate for every model and loop configuration evaluated in RQ1. This split places the aggregate comparison in source context and relates each source to the routing and obligation retention burdens faced by each loop implementation. The aggregate column reproduces the corresponding \textit{w/o} number from Table~\ref{tab:rq1} and equals $\text{RR}_{\text{agg}} = \tfrac{1}{112}\big(29\cdot\text{RR}_{\text{PR}} + 57\cdot\text{RR}_{\text{Course}} + 26\cdot\text{RR}_{\text{Research}}\big)$ up to rounding.
Two findings hold uniformly across model and loop configurations and frame the interpretation of the per source columns.
\textbf{Finding 1: PR Sequences are uniformly the most difficult source.}
Across all 14 model and loop configurations, PR Sequence Resolve Rate reaches at most 3.45\%, and 9 of 14 configurations resolve no PR Sequence tasks. The PR Sequence pool is dominated by merge chains from actively maintained repositories, where each segment often depends on cross module invariants established several PRs earlier. When an early segment fails to preserve a prerequisite invariant, downstream progress becomes governed by obligation retention across dependent edits. This pattern is consistent with the RQ3 observation that regression events remain present across loop profiles, including runs with no recorded internal compaction.
\textbf{Finding 2: Course Labs account for most resolved tasks.}
For every model and loop configuration, Course Labs account for $\geq 70\%$ of the resolved task count. Course Labs have shallower dependency depth on average (median depth $3$ vs.\ $6$ for the pool) and a curriculum imposed module decomposition that often makes the human development order recoverable from task structure. Per source Resolve Rate on Course Labs varies most with model and loop configuration differences, and it tracks the aggregate Resolve Rate ranking most closely.
Research Evolutions fall between the two extremes. The moderate pool size ($26$ tasks) provides a more reliable per source estimate than earlier preview cuts, while the strongest model and loop configurations resolve a limited number of tasks and the weakest configurations resolve none. At this sample size, the per source spread is read together with the column level rate.

\begin{table}[tp]
  \centering
  \caption{Resolve Rate by task source before outer continuation.}
  \label{tab:per-source-rr}
  \small
  \setlength{\tabcolsep}{4pt}
  \renewcommand{\arraystretch}{1.08}
  \begin{tabular}{l l c c c c}
    \toprule
    \textbf{Loop} &  \textbf{Model}
      & \textbf{PR Seq.}\,(n{=}29) & \textbf{Course}\,(n{=}57) & \textbf{Research}\,(n{=}26)
      & \textbf{All}\,(n{=}112) \\
    \midrule
    \multicolumn{6}{@{}p{0.98\linewidth}@{}}{\emph{Model sweep under the Claude Code loop}} \\
    Claude Code & Opus-4.7        & 3.45\%  & 24.56\% & 15.38\% & 16.96\% \\
    Claude Code & GPT-5.5         & 0.00\%  & 21.05\% & 11.54\% & 13.39\% \\
    Claude Code & GLM-5.1         & 0.00\%  & 21.05\% & 11.54\% & 13.39\% \\
    Claude Code & DeepSeek-V4P    & 0.00\%  & 19.30\% & 7.69\%  & 11.61\% \\
    Claude Code & Gemini-3.1-Pro  & 0.00\%  & 14.04\% & 7.69\%  & 8.93\%  \\
    Claude Code & Qwen3.6-Plus    & 0.00\%  & 10.53\% & 3.85\%  & 6.25\%  \\
    Claude Code & Kimi-2.6        & 0.00\%  & 5.26\%  & 3.85\%  & 3.57\%  \\
    Claude Code & Grok-4.1-FR     & 0.00\%  & 5.26\%  & 0.00\%  & 2.68\%  \\
    \midrule
    \multicolumn{6}{@{}p{0.98\linewidth}@{}}{\emph{Loop sweep under a fixed \texttt{gpt-5.4} model}} \\
    Codex          & gpt-5.4 & 0.00\%  & 21.05\% & 11.54\% & 13.39\% \\
    Claude Code    & gpt-5.4 & 0.00\%  & 19.30\% & 11.54\% & 12.50\% \\
    GitHub Copilot & gpt-5.4 & 0.00\%  & 17.54\% & 7.69\%  & 10.71\% \\
    OpenHands      & gpt-5.4 & 0.00\%  & 10.53\% & 3.85\%  & 6.25\%  \\
    SWE-agent      & gpt-5.4 & 0.00\%  & 8.77\%  & 3.85\%  & 5.36\%  \\
    mini-swe-agent & gpt-5.4 & 0.00\%  & 8.77\%  & 0.00\%  & 4.46\%  \\
    \bottomrule
  \end{tabular}
\end{table}

These patterns shape the interpretation of aggregate \lhb{} Resolve Rate. A single aggregate comparison combines Course Labs variation with the limited spread in PR Sequences, and the loop sweep gap in Table~\ref{tab:rq1} is driven largely by Course Labs. PR Sequence gains are clearer in the per source columns than in aggregate deltas. Doubling PR Sequence RR moves the aggregate by less than one percentage point while materially changing the loop engineering interpretation of routing burden and obligation retention.

\FloatBarrier
\section{Context Round and Regression Metrics}\label{app:compaction-internals}

\subsection{Trace Definitions}\label{app:loop-compaction-metrics}

RQ3 counts both external restarts and compaction intervals as context-budget rounds. For trial $i$, let $K_i$ be the number of outer invocations and let $C_i$ be the number of context segments implied by the implementation compaction policy. Each segment opens a new maximal context budget. We report
\begin{equation}
\operatorname{Rounds}=\frac{1}{N_{\mathrm{trace}}}\sum_{i=1}^{N_{\mathrm{trace}}}(K_i+C_i).
\end{equation}
An external restart replaces the active context and is counted in $K_i$. An internal compaction interval is counted in $C_i$. Codex uses the source-level auto-compaction limit, which defaults to $0.9$ of the configured context window. Claude Code uses its configured context budget, and the Ralph SDK analysis uses the SDK compaction interval. Dynamic workflows sum context-budget rounds across the coordinating workflow and worker contexts.

The regression watcher writes \texttt{observation\_history.jsonl}. Let $F_{i,s}$ be the set of released tests that newly fail at triggered snapshot $s$. Regression events per run are
\begin{equation}
\operatorname{Reg/run}=\frac{1}{N_{\mathrm{trace}}}\sum_{i=1}^{N_{\mathrm{trace}}}\sum_s \mathbb{I}[F_{i,s}\neq\varnothing].
\end{equation}
Resolve Rate uses the benchmark outcome for every scheduled task:
\begin{equation}
\operatorname{RR}=\frac{1}{N_{\mathrm{task}}}\sum_{i=1}^{N_{\mathrm{task}}}\mathbb{I}[\text{trial }i\text{ is resolved}].
\end{equation}

\subsection{Run Coverage}

The Codex goal, Claude Code goal, and Ralph runs each contain 102 scheduled tasks. The dynamic workflow run contains 112 scheduled tasks. The corresponding run identifiers are \texttt{stage4-C1-codex-gpt55-0858}, \texttt{cc-cli-opus46-goal-1714}, \texttt{stage4-A1-opus47-1442}, and \texttt{cc-dynamic-opus48-20260705-043454}. Table~\ref{tab:rq3-loop-engineering} uses the available run summaries for Rounds, the available regression histories for Reg/run, and all scheduled tasks for RR.
\subsection{Trace Scope}\label{app:rq3-trace-scope}

The dynamic workflow archive records the coordinating Claude Code session and outer restarts. It does not retain a uniform event stream for every worker process. Its Rounds value therefore describes the top level workflow context. Worker completion state remains visible through repository changes, requirement patches, and regression snapshots.

The four runs use their native model and loop configurations. Their values characterize observed loop profiles and are not a model controlled causal comparison. Missing outer histories are excluded from Rounds and Reg/run, while all scheduled tasks remain in the RR denominator. The figure script reads the archived JSON files directly and applies these rules without manual value entry.
\section{Evaluation Adapter Details}\label{app:adaptations}

\lhb{} evaluates six widely used coding agent loop implementations through a uniform \texttt{BaseAgent} contract. The evaluation adapters normalize credential handling, model identifiers, and workspace conventions, so Table~\ref{tab:rq1} can vary model choice and loop implementation while holding the task contract fixed.

Each run records the evaluation adapter boundary because it determines how an evaluated loop receives requirements, commits work, records plans, and returns artifacts used to compute loop trace metrics.

\paragraph{Workspace and git contract.}

Each evaluated loop executes inside \texttt{/workspace} in a per task Docker container built from the task's \texttt{Dockerfile}, providing every adapter with the same filesystem and execution boundary.

The directory is initialized as a git repository with three subtrees recorded by the evaluation runtime. \texttt{requirements/} contains per unit YAML manifests presented during evaluation, \texttt{agent\_plans/} stores the optional \texttt{plan.json} for RQ2 plan analysis, and \texttt{requirement\_patches/} stores one \texttt{<slug>.diff} per implemented unit derived from \texttt{git diff --cached}.

A unit enters the recorded implementation set after its patch file is non empty. Checkpoint detection (\path{long_horizon_bench/harness/regression_harness.py}) uses \texttt{min\_diff\_lines=5} as the minimum diff growth for a snapshot trigger. The per unit threshold is derived from the smallest gold artifact currently on the ready frontier, making snapshot frequency follow recorded implementation progress rather than transcript length or turn count.

\paragraph{Evaluation adapters.}

\textit{Claude Code} is evaluated through the Anthropic Agent SDK with a \path{ContainerWorkspaceBackend} that proxies the SDK bash and edit tools into the task container. Loop traces and per trial usage are written to \path{claude_sdk_trajectory.jsonl} and \path{claude_sdk_result.json}.

\textit{Codex} uses a custom \texttt{model\_provider} block, \texttt{lhb\_azure} or \texttt{lhb\_proxy}, injected into \texttt{config.toml}. This preserves the Codex invocation contract while routing model calls through the evaluation provider layer, keeping provider transport separate from loop behavior.

\textit{GitHub Copilot CLI} is evaluated in noninteractive mode against a sidecar at \texttt{AZURE\_SIDECAR\_PORT=23337}, reached through the container default gateway IP in default compose setups that omit \texttt{host.docker.internal}.

\textit{Cursor CLI} uses an adapter that maps Anthropic shorthand to Cursor model IDs before invoking \texttt{agent --model}.

\textit{mini-SWE-agent} is installed inside the container at \texttt{/opt/python} through \texttt{pip --prefix} with an explicit \texttt{PYTHONPATH} adjustment. Models prefixed with \texttt{copilot/} are rewritten to \texttt{openai/}, allowing LiteLLM to reach the OpenAI compatible proxy.

\textit{SWE-agent} applies a compatibility patch that removes the \texttt{top\_p} keyword. Anthropic rejects requests carrying both \texttt{temperature} and \texttt{top\_p}, while the SWE-agent defaults set both fields.

\paragraph{Credential isolation.}

Earlier versions of the evaluation adapter wrote Azure AD tokens and Copilot keys into \texttt{os.environ} inside a context manager. Under the \texttt{ThreadPoolExecutor} used by the evaluation runtime, concurrent trials could overwrite shared tokens, and cleanup from one trial could clear a credential entry while another trial remained active. The current evaluation adapters pass credentials as explicit kwargs to \texttt{run\_task} (\texttt{azure\_ad\_token\_provider}, \texttt{copilot\_api\_key}, \texttt{anthropic\_api\_key}), preventing credential races from being attributed to the evaluated loop.
\section{Evaluation Contract for Loop Engineering}\label{app:prompts}

The evaluation contract supplied to each loop implementation is composed by \path{long_horizon_bench/harness/instruction_composer.py}. It combines task wording, planning requirements, and evaluation adapter addenda into the reproducible input observed by the evaluated loop.

For each evaluated loop, the input combines the task instruction, the \emph{Benchmark Contract} (always), the \emph{DAG Plan Requirement} during RQ2 plan fidelity runs, and a profile addendum selected by evaluation adapter class. The Oracle adapter bypasses composition.

The literal prompt text below remains unchanged because it is the evaluated input. The surrounding prose records how that input defines the loop engineering evaluation contract.

\begin{promptbox}{Benchmark Contract \textnormal{(always prepended; \texttt{instruction\_composer.\_BENCHMARK\_CONTRACT})}}
You are running as a long-horizon autonomous engineering agent at `/workspace`.
This task spans many files and requirements, and is expected to take many iterations.

Read every requirement under `/workspace/requirements/` (and any other explicit
requirement list referenced by the task) and only finish once every single one is
addressed. Cherry-picking or partial fixes will lead to failure.

## Optional planning artifact
If you believe the task is complex enough to benefit from a plan, you may write
`/workspace/agent_plans/plan.json` before making changes. This is optional. If you
create it, it must be a JSON object with `nodes` and `edges`. Each node must have
string `id` equal to a requirement slug from `/workspace/requirements/`, may include
`label` and `layer`, and each edge must use string `from` and `to` that reference
those ids.

## Git workflow (required)
`/workspace` is initialized as a git repo. After finishing the implementation for
ONE requirement:
  1. `cd /workspace`
  2. `git add -A`
  3. `git diff --cached > requirement_patches/<slug>.diff`
  4. `git commit -m "impl: <slug>"`
Where `<slug>` is the requirement filename without the `.yaml` extension. The patch
file MUST be non-empty for the requirement to count as complete. Never write or
`touch` an empty `.diff` file.

You may run `git log`, `git diff`, `git show` at any time to inspect what prior
work has done -- only your own commits will be there; no reference solution is
accessible.

## Recommended workflow
1. Read the task carefully and survey `/workspace` so you understand the code you
   are changing.
2. As a development agent, when implementing a requirement you may write tests
   under `/workspace/agent_tests/` to safeguard correctness and future stability --
   write them if you think they help, otherwise skip.
3. Keep iterating until every requirement is implemented. Do not stop at a partial
   fix, a first passing check, or one successful edit.
4. Do not narrow the scope on your own. Missing a requirement is a task failure
   even if the changed tests pass.
5. You may revise the same files and rerun focused checks as many times as needed
   before finishing.
6. Your final summary must report requirement coverage counts, including at least
   the total number of requirements reviewed and how many were implemented and
   blocked.
\end{promptbox}

\begin{promptbox}{Required DAG Plan \textnormal{(replaces the optional planning paragraph for RQ2 plan-fidelity runs)}}
## Required planning artifact (DAG)
Before making any code changes you MUST write `/workspace/agent_plans/plan.json`.
It must be a JSON object with `nodes` and `edges` representing a directed acyclic
graph (DAG) over the requirement set. The DAG should reflect your view of which
requirements can be implemented in parallel and which must be serialized:

- `nodes`: an array. Each node is {"id": "<slug>", "label": "...", "layer": <int>}.
  `id` MUST exactly equal a requirement filename under `/workspace/requirements/`
  without the `.yaml` extension. Every requirement slug MUST appear as a node --
  no omissions.
- `edges`: an array. Each edge is {"from": "<slug>", "to": "<slug>"} meaning the
  `from` requirement must be implemented before `to` (a real ordering constraint,
  e.g. shared file/function, API surface change, or a logical prerequisite). The
  graph MUST be acyclic. Both endpoints MUST exist in `nodes`.
- `layer` is your topological-layer index (0 for sources). Two nodes in the same
  layer should be safely parallelizable -- no edge between them and no implicit
  conflict.

Implement requirements in an order consistent with the DAG. If you discover the
dependency structure is wrong while implementing, update `plan.json` first, then
continue. Do NOT skip writing this file.
\end{promptbox}

\begin{promptbox}{Profile -- \textnormal{\texttt{strong} (Claude Code)}}
Agent profile:
- Focus on completing the task directly.
- Do not spend tokens on basic shell tutorials or extended environment narration.
- Do not silently narrow the task scope, claim success early, or stop after partial
  requirement coverage.
- Use concise progress updates, keep modifying the workspace when needed, and
  validate the parts you change.
\end{promptbox}

\begin{promptbox}{Profile -- \textnormal{\texttt{structured\_open\_source} (SWE-agent, OpenHands)}}
Agent profile:
- Work step by step and keep actions grounded in the repository state.
- Prefer short, concrete progress updates and targeted validation.
- Avoid unnecessary tutorial-style shell explanation.
\end{promptbox}

\begin{promptbox}{Profile -- \textnormal{\texttt{legacy\_verbose} (mini-SWE-agent)}}
Agent profile:
- You may use a slightly more explicit step-by-step style when it helps execution.
- Keep progress updates concrete and repository-specific.
- Prefer direct validation of changed behavior before finishing.
\end{promptbox}

For Codex, GitHub Copilot, and Cursor, the evaluation adapter leaves the profile addendum empty, so the evaluated loop receives the Benchmark Contract and, when applicable, the DAG plan replacement.

\section{Loop Engineering Evaluation Runtime Configuration}
\label{app:run-config}
The \lhb{} evaluation runtime is invoked via \texttt{lhb run --config \textit{config.yaml}}, and the resulting run artifact records the loop parameters used for attribution.
The evaluation runtime treats the evaluation adapter, loop budget, and residual handoff policy as explicit loop engineering factors for each run.
All parameters can be set in the YAML configuration file or overridden by CLI flags. Table~\ref{tab:run-params} summarizes the runtime parameters used to separate attribution across model choice, evaluation adapter, budget, and residual handoff policy while keeping the evaluated task contract fixed.

\begin{table}[!htb]
  \centering
  \caption{Runtime parameters for \texttt{lhb run}.}
  \label{tab:run-params}
  \vspace{0.3em}
  \footnotesize
  \setlength{\tabcolsep}{3pt}
  \renewcommand{\arraystretch}{1.15}
  \resizebox{\columnwidth}{!}{%
  \begin{tabular}{@{}l l l p{6.5cm}@{}}
    \toprule
    \textbf{Group} & \textbf{Parameter} & \textbf{Default} & \textbf{Description} \\
    \midrule
    \multirow{3}{*}{Dataset}
      & \texttt{dataset\_path} & \texttt{tasks/} & Path to the task directory tree. \\
      & \texttt{task\_ids} & all               & Task ID globs to include (repeatable). \\
      & \texttt{exclude\_task\_ids} & \texttt{[]} & Task ID globs to exclude (repeatable). \\
    \midrule
    \multirow{4}{*}{Evaluation adapter}
      & \texttt{agent} & \texttt{oracle} & Built in evaluation adapter name. \\
      & \texttt{agent\_import\_path} & --             & Import path for a custom evaluation adapter class. \\
      & \texttt{model} & --                & LLM in \texttt{provider/model} format. \\
      & \texttt{agent\_kwargs} & \texttt{\{\}} & Key value pairs forwarded to the evaluation adapter. \\
    \midrule
    \multirow{3}{*}{Output}
      & \texttt{output\_path} & \texttt{runs/} & Root directory for run artifacts. \\
      & \texttt{run\_id} & timestamp         & Unique identifier for this run. \\
      & \texttt{n\_tasks} & all               & Cap on number of tasks to evaluate. \\
    \midrule
    \multirow{2}{*}{Build}
      & \texttt{no\_rebuild} & \texttt{false} & Skip rebuilding Docker images. \\
      & \texttt{cleanup} & \texttt{true} & Remove images after run completes. \\
    \midrule
    \multirow{2}{*}{Logging}
      & \texttt{log\_level} & \texttt{info} & Python logging level. \\
      & \texttt{livestream} & \texttt{false} & Stream evaluated loop I/O in real time. \\
    \midrule
    \multirow{2}{*}{Concurrency}
      & \texttt{n\_concurrent} & \texttt{4} & Tasks evaluated in parallel. \\
      & \texttt{n\_attempts} & \texttt{1} & Attempts per task. \\
    \midrule
    \multirow{3}{*}{Timeout}
      & \texttt{global\_timeout\_multiplier} & \texttt{1.0} & Multiplier for all per task timeouts. \\
      & \texttt{global\_agent\_timeout\_sec} & per task     & Override loop budget (seconds). \\
      & \texttt{global\_test\_timeout\_sec} & per task     & Override test budget in seconds. \\
    \bottomrule
  \end{tabular}}
\end{table}

\paragraph{Per task timeouts.}
Each task declares \texttt{max\_agent\_timeout\_sec} (default~1\,800\,s) and \texttt{max\_test\_timeout\_sec} (default~3\,600\,s) in its \texttt{task.yaml}.
The global overrides and multiplier allow uniform budget scaling without rewriting individual task files. Budget changes therefore remain attributable to run configuration rather than task materialization.

\paragraph{Residual handoff structure.}
The evaluation runtime executes an \emph{outer loop} over evaluation adapter runs. Each round starts a fresh shell session while preserving the workspace from prior rounds, so residual handoff is measured as an observed loop engineering factor rather than treated as an implicit retry mechanism.
The number of rounds is controlled by \texttt{outer\_loop\_count} (default~3); setting it to \texttt{0} or \texttt{unlimited} removes the cap, allowing repeated residual handoffs until the timeout budget is exhausted and making continuation policy explicit in the loop trace.
Within each round, rate limit errors trigger retries governed by \texttt{rate\_limit\_retries\_per\_round} (default~5) with exponential backoff starting at \texttt{rate\_limit\_backoff\_sec} (default~45\,s), recording provider availability separately from loop progress.

\paragraph{Model and provider configuration.}
Provider credentials and environment variables are supplied via a separate YAML file (\texttt{model\_config\_path} or \texttt{-{}-model-config}), which supports \texttt{env\_files} (dotenv paths) and an explicit \texttt{env} map. This separation places provider transport outside the loop engineering factors attributed to model choice and evaluation adapter.
Keys already set in the process environment are preserved unless \texttt{override\_process\_env:\allowbreak{} true} is selected, reducing cross run configuration drift.

\paragraph{Example configuration.}
The following minimal configuration specifies the dataset and execution policy while recording the evaluation adapter, loop budget, and residual handoff policy as explicit run metadata.

\begin{promptbox}{Run Configuration \textnormal{(\texttt{lhb.run.yaml})}}
dataset_path: tasks
output_path: runs

agent: mini-swe-agent
model: anthropic/claude-sonnet-4-20250514

n_concurrent: 4
n_attempts: 1

no_rebuild: false
cleanup: true
log_level: info
livestream: false

agent_kwargs:
  outer_loop_count: 3
  rate_limit_retries_per_round: 5
  rate_limit_backoff_sec: 45

global_timeout_multiplier: 1.0
\end{promptbox}

\begin{promptbox}{Model Environment \textnormal{(\texttt{lhb.model.yaml})}}
env_files: []
override_process_env: false
env:
  ANTHROPIC_API_KEY: "sk-ant-..."
\end{promptbox}

\end{document}